\documentclass[superscriptaddress,twocolumn,secnumarabic,amssymb,nobibnotes,aps,prb]{revtex4-2}
\usepackage{siunitx}
\usepackage{graphicx}
\usepackage{amsmath}
\usepackage{amsfonts}
\usepackage{xcolor}

\def\Tg{$T_{\text{g}} $}
\def\Tstar{$T^* $}
\def\Tgstar{$T^*_\text{g} $}

\begin{document}
\title{The protein dynamical transition is independent of hydration}

\author{Johanna K\"{o}lbel}
\affiliation{Department of Chemical Engineering, University of Cambridge, Cambridge, CB3 0AS, UK}
\affiliation{Current affiliation: School of Engineering, Brown University, 184 Hope Street, Providence, RI 02912, USA}

\author{Moritz~L.~Anuschek}
\affiliation{Department of Pharmacy -- Center for Drug Research, Pharmaceutical Technology and Biopharmaceutics, Butenandtstrasse 5, 81377 Munich, Germany}
\affiliation{Current affiliation: Novo Nordisk, Novo Nordisk Park 1, 2760 Måløv, Denmark}

\author{Ivonne~Stelzl}
\affiliation{Department of Pharmacy -- Center for Drug Research, Pharmaceutical Technology and Biopharmaceutics, Butenandtstrasse 5, 81377 Munich, Germany}
\affiliation{Current affiliation: Coriolis Pharma Research GmbH, Fraunhoferstraße 18b, 82152 Martinsried, Germany}

\author{Supawan~Santitewagun}
\affiliation{Department of Chemical Engineering, University of Cambridge, Cambridge, CB3 0AS, UK}

\author{Wolfgang~Frie\ss}
\affiliation{Department of Pharmacy -- Center for Drug Research, Pharmaceutical Technology and Biopharmaceutics, Butenandtstrasse 5, 81377 Munich, Germany}

\author{J. Axel Zeitler}
\email{jaz22@cam.ac.uk}
\affiliation{Department of Chemical Engineering, University of Cambridge, Cambridge, CB3 0AS, UK}

\date{\today}

\begin{abstract}
Terahertz time-domain spectroscopy and differential scanning calorimetry were used to study the role of the dynamics of biomolecules decoupled from solvent effects. Lyophilised sucrose exhibited steadily increasing absorption with temperature as anharmonic excitations commence as the system emerges from a deep minimum of the potential energy landscape where harmonic vibrations dominate. The polypeptide bacitracin and two globular proteins, lysozyme and human serum albumin, showed a more complex temperature dependence. Further analysis focused on the spectral signature below and above the boson peak. We found evidence for the onset of anharmonic motions that are characteristic for partial unfolding and molecular jamming in the dry biomolecules. The activation of modes of the protein molecules at temperatures comparable to the protein dynamical transition temperature was observed in the absence of hydration. No evidence for Fr{\"o}hlich coherence, postulated to facilitate biological function, was found in our experiments. 
\end{abstract}
\maketitle

\section{Introduction}
The number of protein drugs in therapy is steadily increasing. These biopharmaceuticals typically need to be administered by injection. Due to the intrinsically limited stability of the molecules in aqueous solution, the protein-drug products are frequently freeze-dried into an amorphous solid matrix for storage and reconstituted immediately before use.

The molecules surrounding the protein molecules in solution constitute their solvation shell (sometimes also called hydration shell); and the molecular mobility of this shell affects the rates of conformational change, catalysis, and protein/DNA-protein interactions \cite{aoki2016salt, dahanayake2018does}. 
The misfolding propensity and the pathways of protein aggregation depend on the protein's local environment, which is influenced by the solvent, water, as well as sugars, salts, metal ions, and lipids that are part of the physiological environment or the reconstituted formulation. Molecular crowding at high protein concentration, pH and buffer also play a role \cite{roeters2017evidence, galvagnion2015lipid, munishkina2008concerted, binolfi2006interaction, uversky2001evidence, giehm2010strategies}. The predominant intermolecular interactions that proteins form with their surrounding environment are hydrogen bonds with water.

A recent paper highlighted the importance that the   mobility of water plays in protein dynamics and, ultimately, aggregation. Simulations in combination with various experimental techniques focusing on the intrinsically disordered model protein $\alpha$-synuclein (aSyn) showed that water mobility and aSyn mobility are inextricably linked. Enhancing the water mobility reduces the propensity of aSyn to aggregate \cite{stephens2022decreased}.

The timescales of solvent motions and conformational changes in proteins differ significantly. Solvent motions are rapid, occurring on the femtosecond to picosecond timescale, whereas conformational changes in proteins happen on the nanosecond to millisecond timescale. Still, solvent mobility affects protein motions \cite{khodadadi2015protein}.
The coupling of water motions, the presence of ions, and the protein dynamics are protein specific due to differing charge distribution, hydrophobicity, and surface roughness \cite{janc2018ion}.

Thus, protein unfolding in solution occurs via a complex pathway and the properties of the hydration shell are of critical importance \cite{leitner2008solvation}. These interactions are restrained in lyophilised samples with strongly reduced water content. Conversely, complete unfolding in the solid state is usually not observed as thermal decomposition occurs before sufficiently high temperatures for unfolding are reached.

Instead of studying solvated biomolecules, here we removed all possible water molecules from protein samples. This results in the molecules coming into close contact but without aggregating, as was the subject of a previous study by Stephens et al. \cite{stephens2022decreased}. We can thus study protein dynamics decoupled from solvent effects. This is of fundamental interest, as well as of practical importance, for lyophilisates of protein drugs \cite{tang2004design}.

Lyophilisates typically comprise cryo- and lyoprotectants, surfactants, and the active biomolecule. The design of the products and the lyophilisation process require good understanding of the underlying stabilisation mechanisms of the formulation components \cite{cicerone2015stabilization, tang2004design}.

In this context, the residual water content in dried protein samples plays an essential role at low temperatures. Lyophilised lysozyme samples with residual hydration larger than \SI{27}{\percent (m/m)} water content \cite{knab2006hydration}, well above the typical water content of lyophilisates of approximately \SI{1}{\percent (w/w)} exhibit the so-called protein dynamical transition (PDT), an increase in the mean square displacement of molecules at a temperature of \SIrange{180}{220}{K} upon heating. 
The PDT that is usually measured with neutron scattering is not to be confused with the glass transition \Tg\ that occurs in amorphous samples, which is characterised by a sudden change in relaxation times and heat capacity  at \Tg\ and is usually measured with DSC. \cite{sibik2013glassy}
The PDT has been observed at temperatures at around \SI{200}{K} for different lightly hydrated proteins and is thought to be due to the onset of motions involving interactions of charged side chains with surrounding water molecules. \cite{he2008protein}

The PDT and the glass transition can both be described in terms of Goldstein's potential energy landscape (or surface, PES) containing many small minima within larger basins \cite{cavagna2009supercooled, goldstein1969viscous}. Moving between minima requires overcoming the local, shallow activation energy barriers. With lower temperature, the configurational entropy decreases, i.e. the number of available minima decreases. Moving from one basin to the next requires a large amount of activation energy and cooperative rearrangement of molecules. The protein dynamical and the glass transitions can be understood as corresponding to lowering energy barriers on that landscape. 
In the absence of hydration, a temperature increase activates a different set of motions of activation energies similar to the thermal energies supplied. This effect can be very subtle.

Given the propensity of hydrogen and van der Waals bonding interactions in solvated and lyophilised biopharmaceuticals, an ideal experimental method to study such systems is terahertz time-domain spectroscopy, THz-TDS, due to the match in photon and bonding energies \cite{shmool2019insights}. In the terahertz range, the boson peak (BP) and the coupling of dipoles to the vibrational density of states dominate the absorption mechanisms \cite{kolbel2022terahertz}.
The motions at terahertz frequencies play an essential role in understanding solid-state protein dynamics. At storage temperature (room temperature), a considerable number of vibrational modes in the terahertz frequency range are active and contribute to the formulation stability.

In the 1970s, Fr{\"o}hlich suggested the existence of coherent vibrations at around \SI{e11}{Hz} and postulated that such motions enable biological functions, so-called biological control through selective long-range interactions. Motions active at terahertz frequencies are long-range motions, and the possible existence of a so-called Fr{\"o}hlich condensate in lyophilised protein formulations can hence be investigated. Thus far no experimental evidence  for such coherent states was found for biomolecules in solution.  

Vibrational confinement can dramatically reduce molecular mobility in lyophilisates at temperatures close to room temperature and depends heavily on the interaction between protein and excipients in the formulation \cite{shmool2019observation}. Upon increasing the temperature, the molecular mobility in the sample increases due to the ability to access more of the local minimum in the PES until the free volume is taken up entirely and the molecule becomes ``jammed''. Any further increase in mobility, and hence terahertz absorption, is no longer possible until a higher energy barrier to another basin in the PES is overcome that is associated with additional degrees of freedom for molecular motions.
 
In the spectral region accessible with most terahertz spectrometers, between \SI{0.3}{THz} and \SI{3.0}{THz}, the frequency-dependent infrared absorption coefficient $\alpha(\omega)$ is theoretically related to the reduced density of states $g(\omega)$ via $\frac{g(\omega)}{\omega^2}\propto\frac{\alpha(\omega)}{\omega^3}$ \cite{kolbel2022terahertz}.
The excess density of states, apparent in the reduced density of states, is referred to as the boson peak (BP). The BP is a harmonic phenomenon due to inherent disorder that anharmonic effects can obscure \cite{schirmacher1998harmonic}.
Utilising THz-TDS, the onset temperature of molecular mobility was found to correlate with anharmonic effects in the model glass-former glycerol. These anharmonic effects resulted in an apparent shift of the BP centre frequency and could thus be separated from purely harmonic contributions \cite{kolbel2022terahertz}.

Markelz et al. \cite{markelz2007protein} observed that in an amorphous sample, an increase in absorption with temperature at a single frequency as measured with THz-TDS is due to anharmonic effects, even at very low temperatures. While no actual PES is perfectly harmonic, these effects are comparatively small at low temperatures, e.g. the BP is not yet obscured or its apparent centre frequency affected. We will hence refer to the temperature region below \Tstar, i.e. at temperatures at which the BP is unaffected by anharmonic effects, as the harmonic regime, and to the temperature region at which the BP is affected as the anharmonic regime.
This is expected to conceptually also apply to protein samples. 

\begin{figure}
\centering
      \includegraphics[angle=90,width=0.49\linewidth]{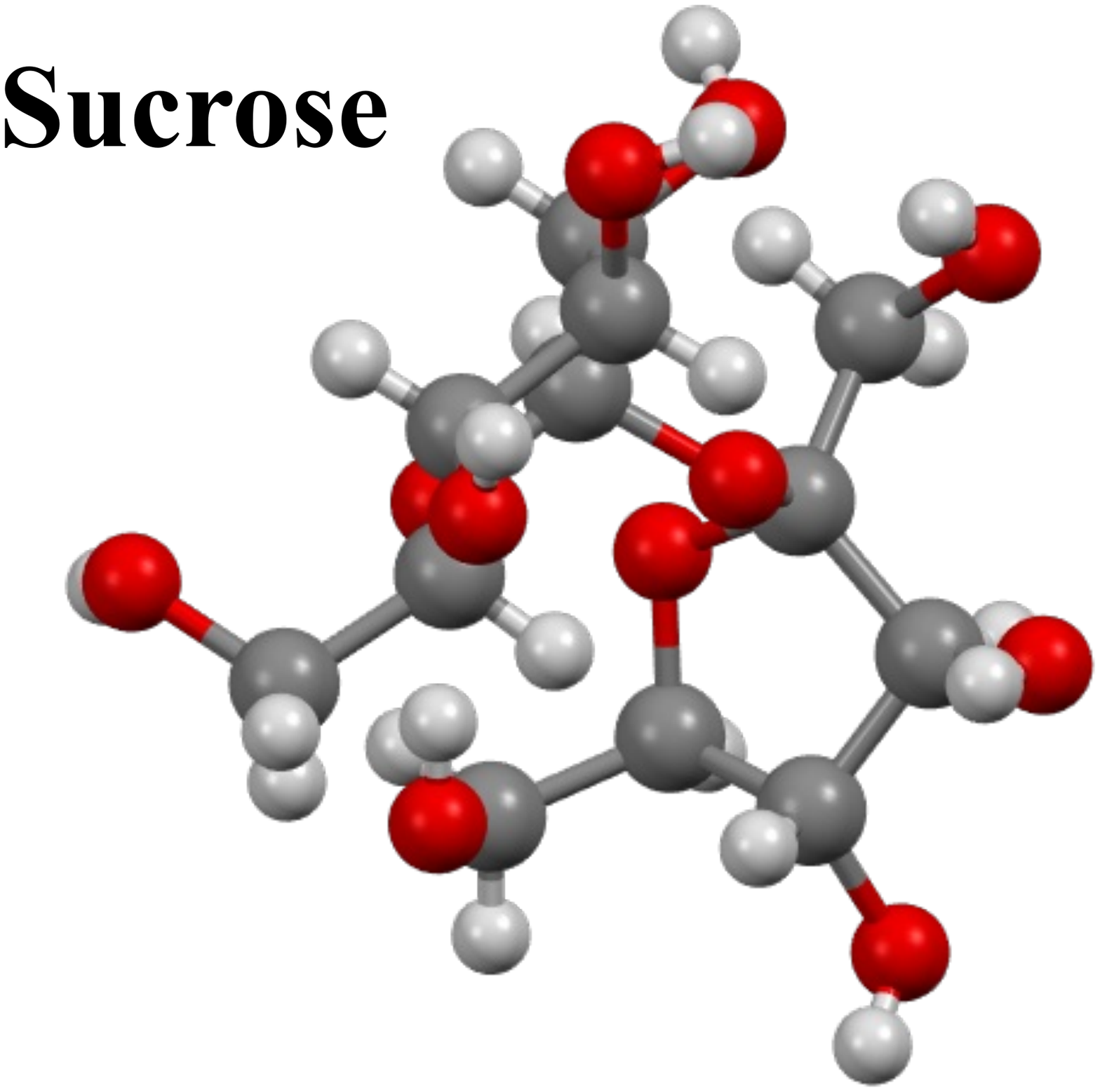}      \includegraphics[angle=90,width=0.49\linewidth]{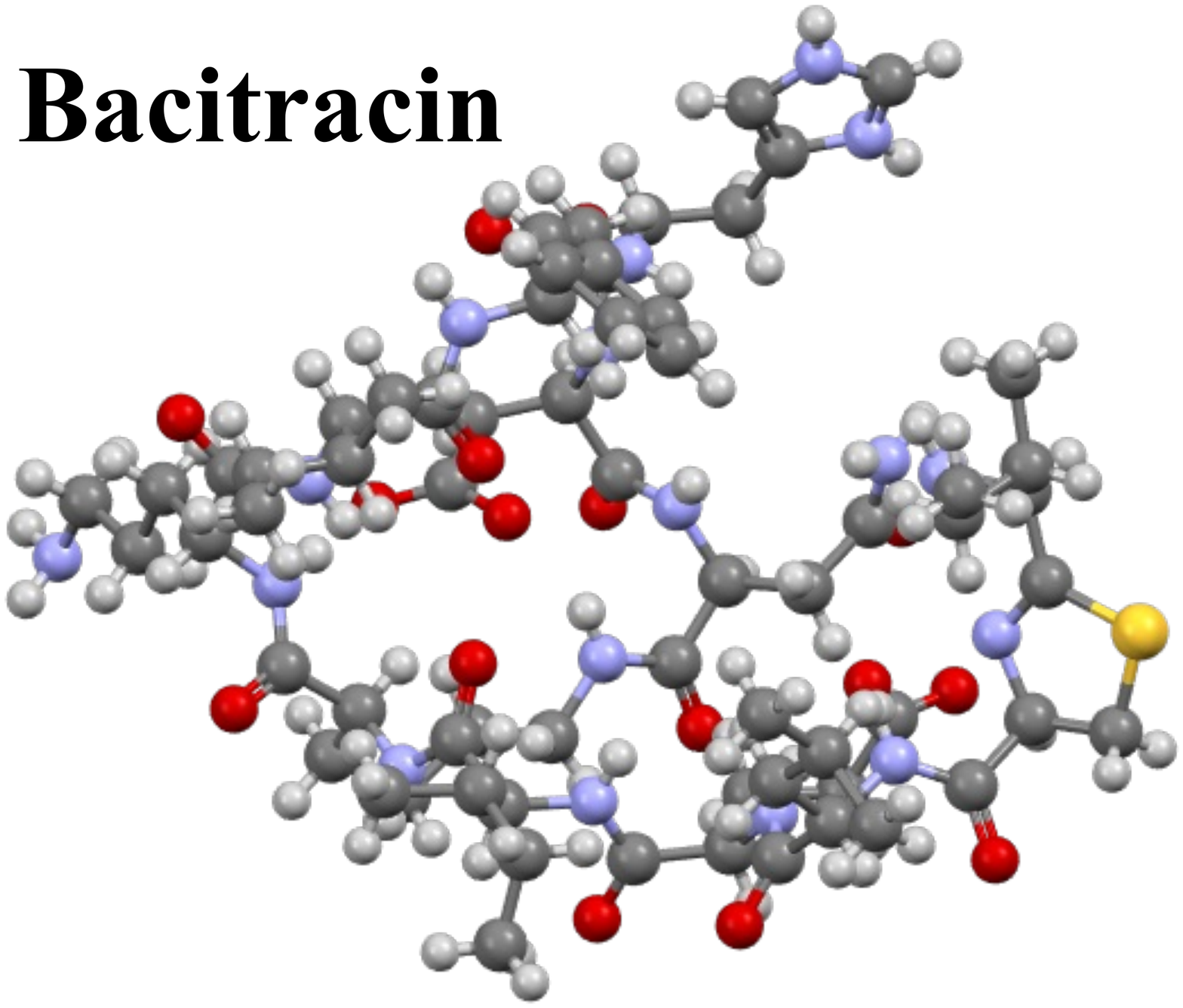}      \includegraphics[angle=90, width=0.49\linewidth]{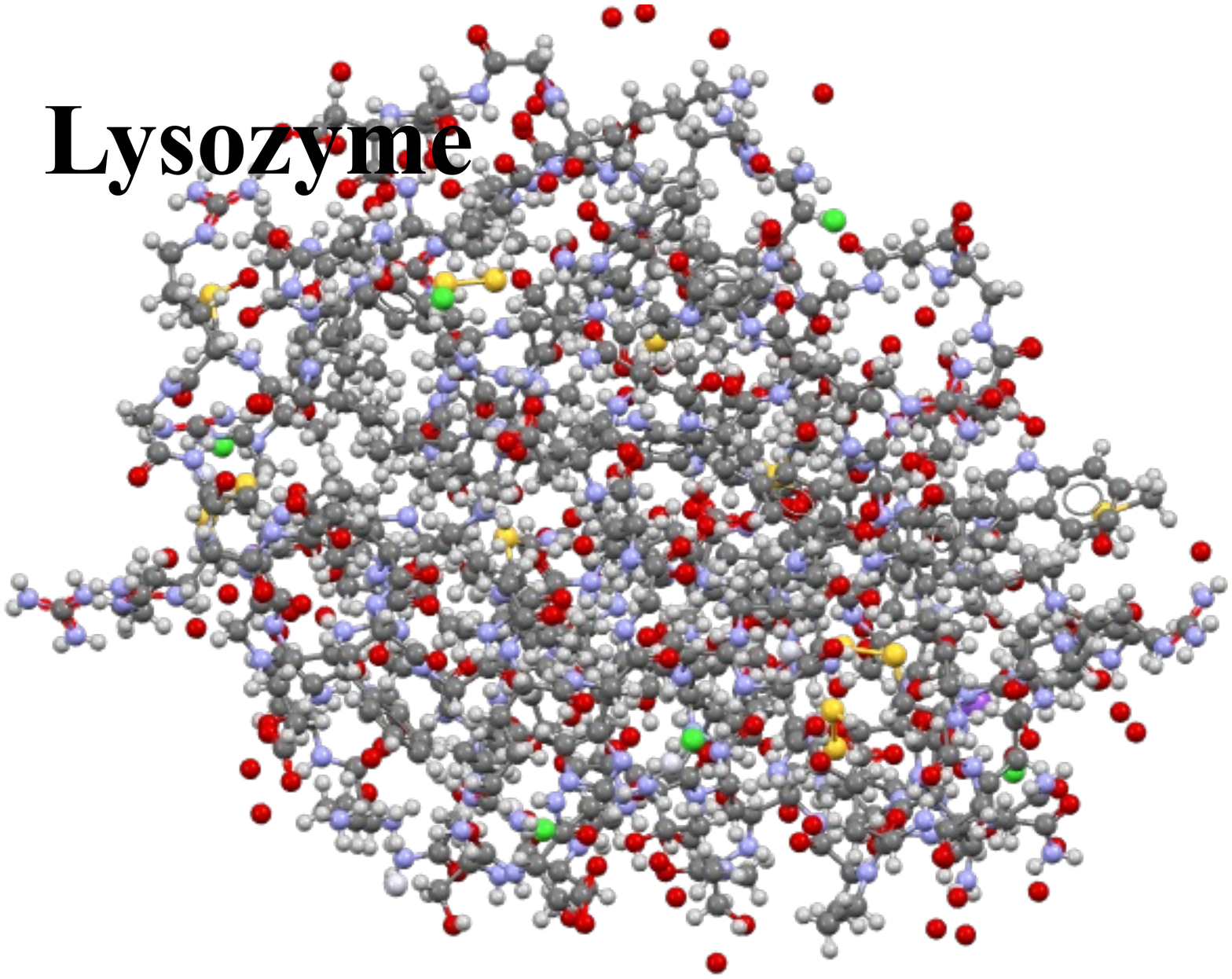}
      \includegraphics[angle=90,width=0.49\linewidth]{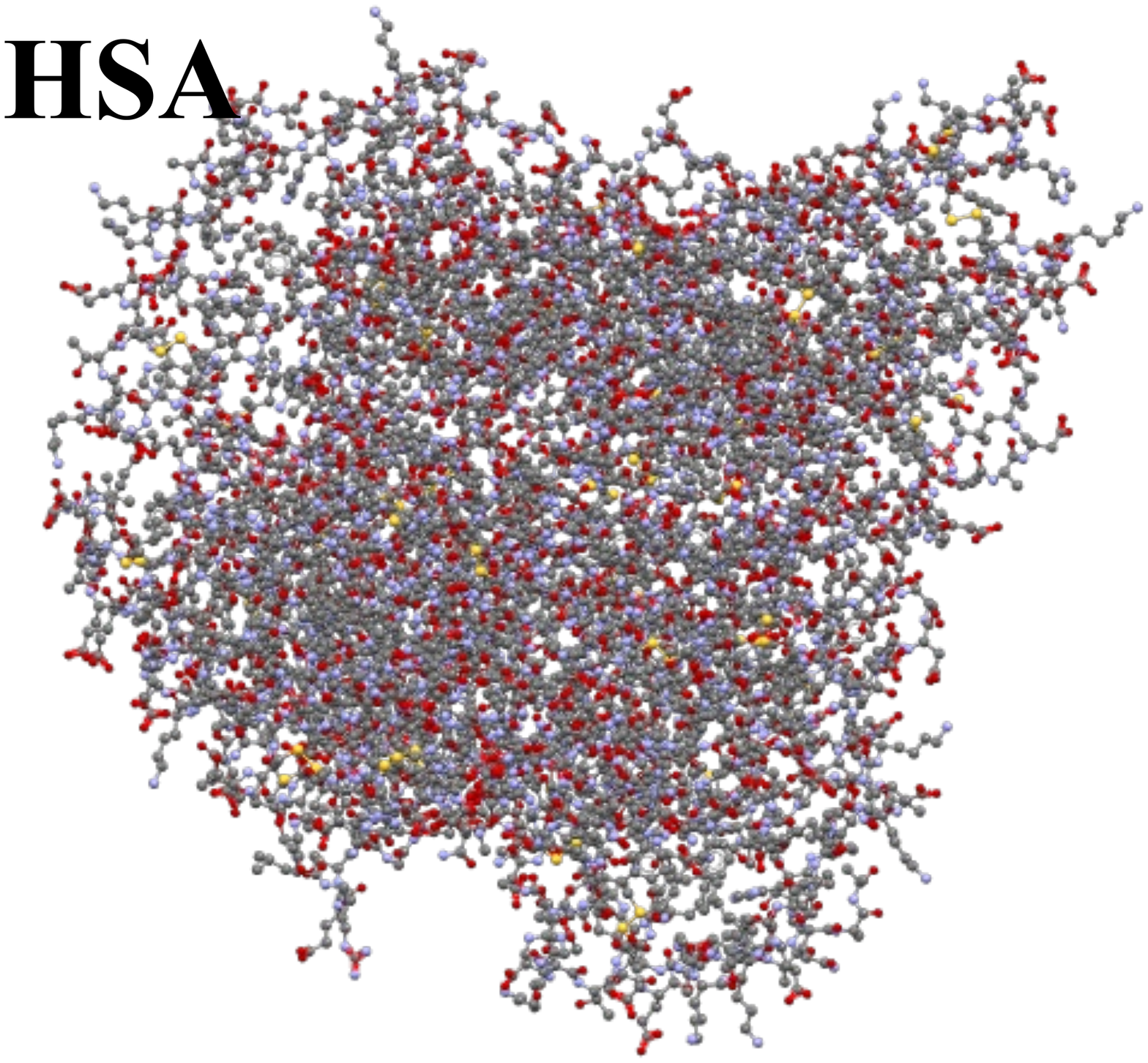}
	\caption{Visualisation of the four different key molecules of this study. \label{fig:cif}}
\end{figure}

In the present work, we investigated terahertz protein dynamics decoupled from solvent effects in four different one-component lyophilised products, namely sucrose (a widely used bulking agent, molecular weight \SI{0.34}{kDa}), bacitracin (a polypeptide antibiotic, molecular weight \SI{1.4}{kDa}), lysozyme (a globular protein, molecular weight \SI{14.5}{kDa}), and human serum albumin (HSA, a globular protein and bulking agent, molecular weight \SI{66.5}{kDa}), which are shown in Figure \ref{fig:cif}. A possible Fr{\"o}hlich condensate in lyophilised protein formulations would become apparent by distinct spectral features emerging in the terahertz spectrum.

\section{Materials and Methods}
\subsection*{Materials}
Sucrose (a commonly used cryo- and lyoprotectant, molecular weight \SI{0.34}{kDa}) and HSA (globular model protein, molecular weight \SI{66.5}{kDa}) were purchased from Merck GmbH (Steinheim, Germany). Bacitracin (a polypeptide antibiotic, molecular weight \SI{1.4}{kDa}) and lysozyme (a globular protein, molecular weight \SI{14.5}{kDa}) were purchased from Carl Roth GmbH (Karlsruhe, Germany). 
The samples were prepared with highly purified water (HPW; Sartorius Arium Pro, Sartorius, G\"{o}ttingen, Germany) to reach a total solid content of \SI{10}{\percent m/m} prior to lyophilisation.

\subsection*{Lyophilisation}
Lyophilisation stoppers (B2-TR coating, West) and DIN 10R vials (Fiolax, Schott, Germany) were cleaned with highly purified water and dried at \SI{333}{K} for \SI{8}{h}. The vials were filled with \SI{3}{mL} solution and subsequently semi-stoppered. The product temperature in vials at different positions on the shelf was recorded with a thermocouple. Formulations were freeze-dried according to the protocol in Table~\ref{tab:lyo} using an FTS LyoStar 3 freeze dryer (SP Scientific, Warminster, Pennsylvania, USA). The end of primary drying was controlled by comparative pressure measurement between a Pirani and MKS sensor. The vials were stoppered after secondary drying under nitrogen atmosphere at \SI{800}{mbar} and crimped with flip-off seals. 

\begin{table}\caption{Lyophilisation protocol.}\label{tab:lyo}\centering
\resizebox{\linewidth}{!}{
\begin{tabular}{lcccc}
\hline
  Step&Ramp (\SI{}{K\per\min})& Shelf temperature (\SI{}{K})& Pressure (\SI{}{\micro bar}) & Hold time (\SI{}{h}) \\
  \hline
  Freezing & 1.0&	223&	1\,atm&	3 \\
   Primary drying& 0.5&	253&	60&	50 \\
      Secondary drying& 0.4&	323&	60&	5 \\
  \hline
\end{tabular}}
\end{table}

\subsection*{Differential Scanning Calorimetry (DSC)}
The \Tg\ of the lyophilisates was determined with a DSC 821$^\text{e}$ (Mettler Toledo, Giessen, Germany). 5 to \SI{10}{mg} of crushed lyo cake were filled into \SI{40}{\micro\liter} aluminium crucibles (Mettler Toledo, Giessen, Germany) under controlled humidity conditions ($\leq$\SI{10}{\percent} relative humidity) and sealed hermetically. The samples were heated from \SI{280}{K} to \SI{415}{K} at a ramp rate of \SI{2}{K min^{-1}}. The \Tg\ was determined as the midpoint of the phase transition.

\subsection*{Terahertz Time-Domain Spectroscopy (THz-TDS)}
 
The lyophilised cake was broken up under a dry nitrogen atmosphere contained within a glove bag (AtmosBag, Merck UK, Gillingham, UK), the powder was gently mixed using an agate mortar and pestle and then pressed into a thin pellet (thickness less than \SI{800}{\micro\meter}, diameter \SI{13}{\milli\meter}) using a manual press (load \SI{3}{\tonne}, Specac Ltd, Orpington, UK). The pellet was  sealed between two z-cut quartz windows of \SI{2}{mm} thickness each and fixed to the cold finger of a cryostat (ST-100, Janis, Wilmington, MA, USA). 

Samples were analysed with a Terapulse 4000 spectrometer (Teraview Ltd, Cambridge, UK) in transmission  under vacuum (pressure $<\SI{20}{\milli\bar}$). Each sample and reference spectrum was calculated from the co-average of 1000\,waveforms which were acquired with a resolution of \SI{0.94}{\per\centi\meter} and transformed to the frequency domain via fast Fourier transform. 
The absorption coefficient was calculated following the method by Duvillaret et al. \cite{duvillaret1996reliable}

At the beginning of each measurement, the sample was cooled down from room temperature to \SI{80}{\kelvin} and left to equilibrate for at least \SI{30}{\minute}. The temperature was subsequently increased in steps of \SI{10}{\kelvin} up to a maximum temperature of  \SI{440}{\kelvin}. The system was allowed to equilibrate for \SI{8}{\minute} at each temperature increment before  reference (two z-cut quartz windows with no sample in between) and sample measurement.

\section{Results and discussion}

\subsection{Thermal analysis of \Tg\ by DSC}

\begin{figure}
\centering
      \includegraphics[width=1\linewidth]{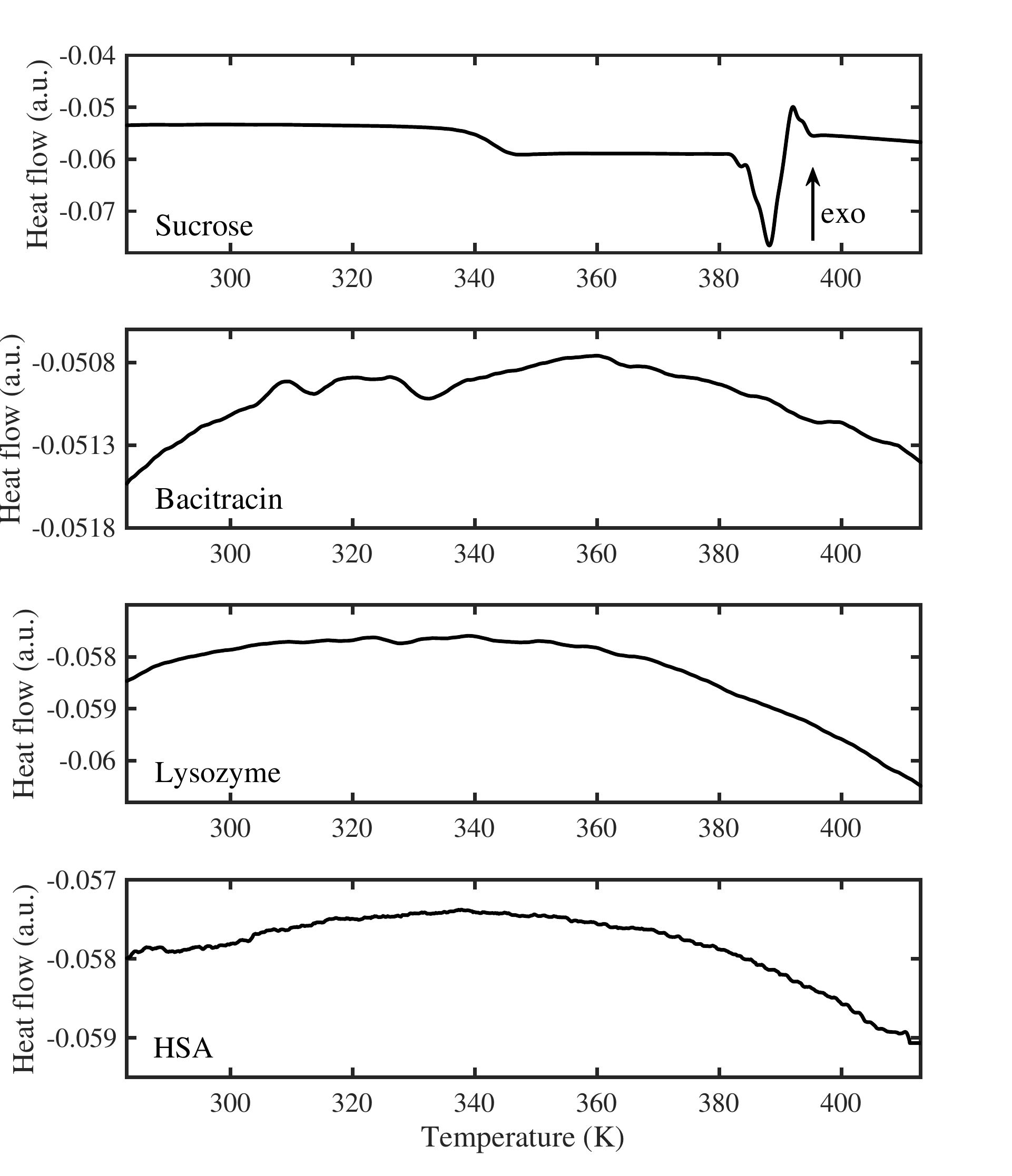}
	\caption{DSC of sucrose, bacitracin, lysozyme, and HSA lyophilisates. A clear \Tg\ was only found for sucrose. Inflection points for the other samples were found at \SI{348}{K} (bacitracin), \SI{330}{K} (lysozyme), and \SI{337}{K} (HSA).  
	\label{fig:dsc}}
\end{figure}

In DSC data a clear \Tg\ was only found for sucrose (at \SI{340}{K}, shown in Figure \ref{fig:dsc}). The peptide and proteins did not show a clear step in heat capacity corresponding to \Tg; instead, a gradual decrease in heat flow was observed, potentially linked to structural changes at elevated temperatures. The time scales of protein unfolding are strongly temperature-dependent and it is possible that the mobility was insufficient to maintain equilibrium between folded and unfolded states during the DSC measurements  \cite{chang2009mechanisms}. In all three pure macromolecule samples, an inflection point occurred, namely at \SI{384}{K} (bacitracin), \SI{330}{K} (lysozyme), and \SI{337}{K} (HSA).

\subsection{THz-TDS provides insights into molecular mobility}

\begin{figure}
\centering
      \includegraphics[width=1.1\linewidth]{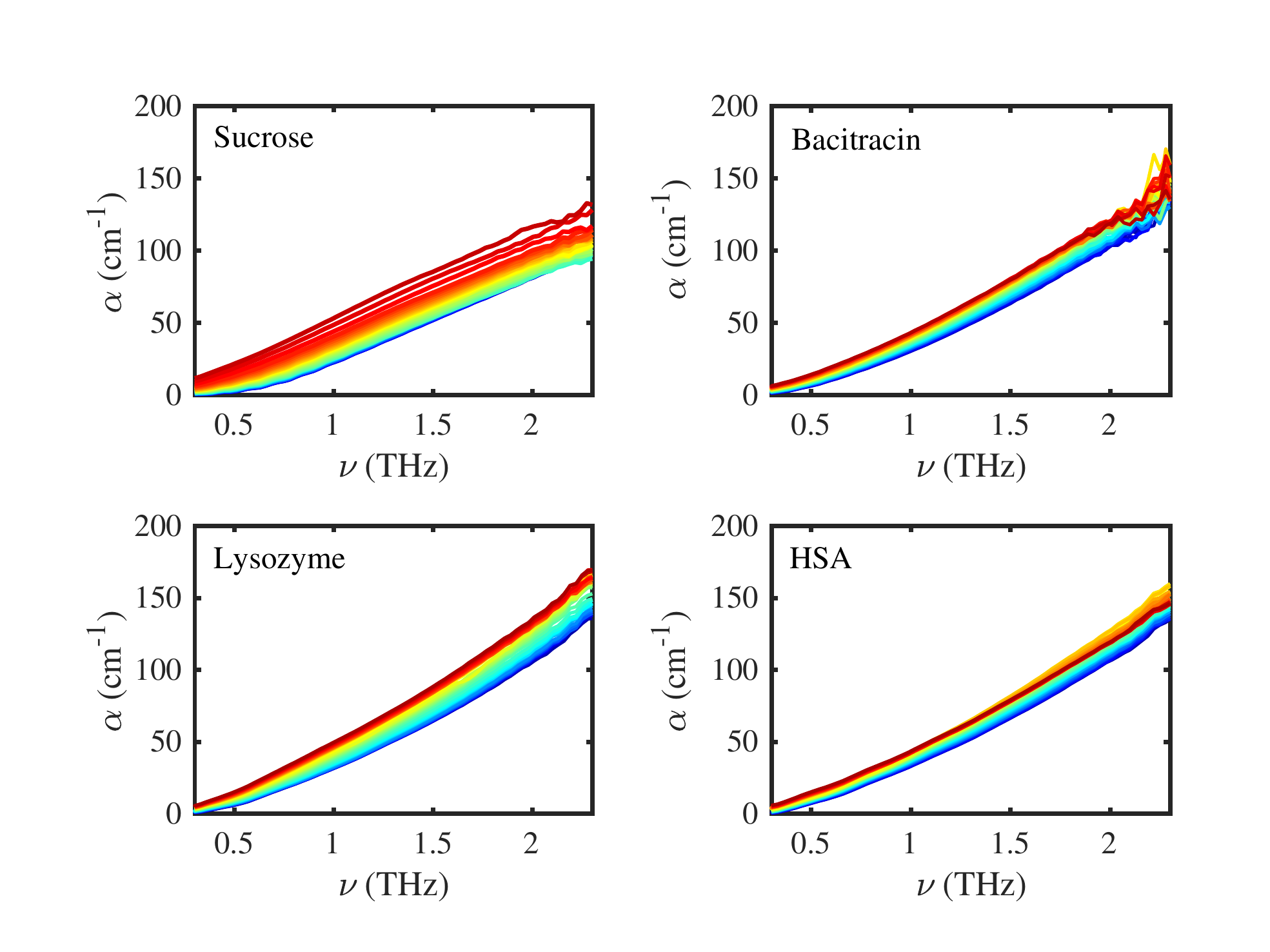}
	\caption{Terahertz spectra for sucrose, bacitracin, lysozyme, and HSA lyophilisates. Absorption mostly increases with temperature. Blue: \SI{80}{\kelvin}, red: \SI{420}{\kelvin}. \label{fig:spectra}}
\end{figure}

\begin{figure}
\centering
       \includegraphics[width=\linewidth]{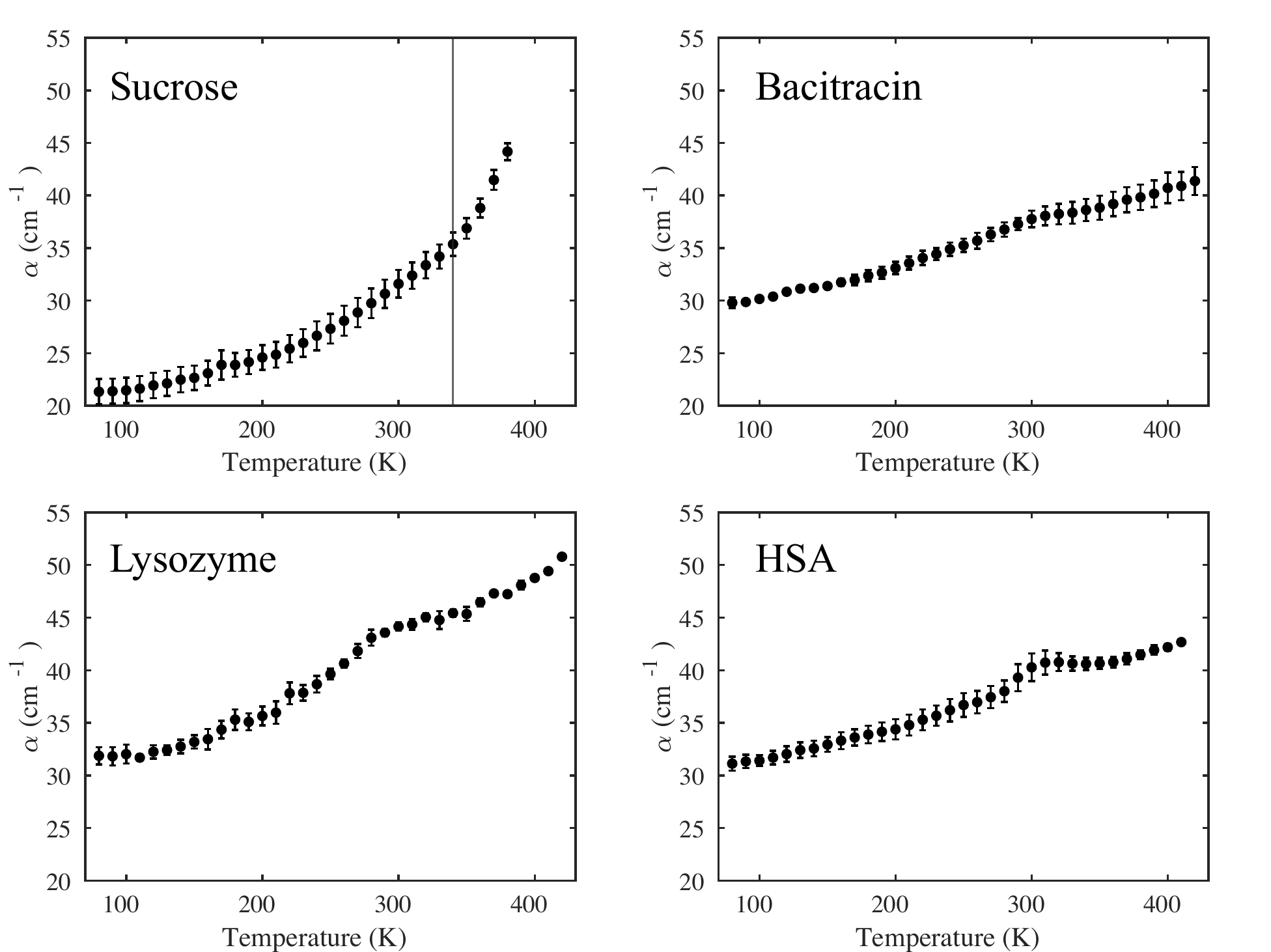}
	\caption{Absorption at \SI{1}{\tera\hertz} for sucrose, bacitracin, lysozyme, and HSA lyophilisates. The vertical line in the plot of sucrose marks the \Tg\ as determined with DSC. Error bars are standard error for $n$ measurements ($n=5$ for sucrose, $n=4$ for bacitracin, $n=3$ for lysozyme and HSA). \label{fig:absat1_stacked}}
\end{figure}

The THz-TDS spectra of the four materials show a similar profile (Figure \ref{fig:spectra}). The change in absorption coefficient with temperature was most pronounced for sucrose. 
Figure \ref{fig:absat1_stacked} shows the absorption coefficient extracted at a frequency of \SI{1}{THz} for different samples. The absorption coefficient measures the change in dipole moments caused by inter- and intramolecular motions in the sample of interest. 
As the temperature increases, larger-scale motions become available, resulting in changes of the dipole moments. Each sample is characterised by a distribution of states, each with its own onset temperature. In smaller systems, like sucrose, the distribution of states is narrower in temperature, and the total number of states is lower than in more complex systems.

The rate of change in absorption with temperature increased at both transition temperatures (see Figure \ref{fig:absat1_stacked}). 
\Tgstar\ was found at \SI{340}{\kelvin}, which agreed very well with the \Tg\ measured by DSC and literature values \cite{roe2005glass}. \Tstar, which cannot be measured by DSC, was found at \SI{230}{\kelvin}.

In contrast, the changes in the absorption coefficient of the protein samples can be very gradual with temperature. The PDT commonly occurs around \SI{200}{K}. We therefore analysed the average rate of absorption change in three temperature intervals: $T < \SI{200}{K}$, $ \SI{200}{K}< T < \SI{300}{K}$, and $T> \SI{300}{K}$ (Table \ref{tab:gradient}). 

\begin{table}
\caption{Rate of absorption change with temperature of sucrose, bacitracin, lysozyme, and HSA lyophilisates. 
}
\resizebox{\linewidth}{!}{
\begin{tabular}{l|c|c|c|c}
& \multicolumn{4}{c}{$\text{d}\alpha/\text{d}T$ (\SI{e-2}{\per\centi\meter\per\kelvin})}  \\
\hline
  &Sucrose& Bacitracin& Lysozyme & HSA \\
  \hline
 $T <\SI{200}{\kelvin}$ &	$2.9 \pm 0.4$ 	&$2.7 \pm 0.2$&	$3.5 \pm 0.8$	&$2.9 \pm 0.2$\\
 $\SI{200}{\kelvin}<T<\SI{300}{\kelvin} $ &$7.1 \pm 0.6$	&$4.6 \pm 0.2$	&$9.0 \pm 0.8$	&$5.5 \pm 0.8$\\
 $T>\SI{300}{\kelvin}$ 	&$15.2 \pm 3.3$	&$3.0 \pm 0.3$&	$5.3 \pm 0.8$	&$1.9 \pm 0.6$\\ 
\end{tabular}}
\label{tab:gradient}
\end{table}

It is known that the \Tg\ of sucrose decreases with increasing water content \cite{roe2005glass}. The agreement between the \Tgstar\ we measured with THz-TDS as well as DSC with literature values for dry sucrose matrices indicates that the experimental setup and procedures ensure a very  low water content. 
Additionally in THz-TDS, any water molecules that could have potentially adsorbed to the sample surface during  preparation may be removed by the vacuum (of $<$\SI{20}{mbar}) in the measurement chamber.

Generally, the rate of change in absorption with temperature decreased at temperatures above \SI{300}{\kelvin} for the larger molecular weight systems. 
This phenomenon was previously observed in other (more complex) lyophilised formulations and attributed to high-temperature macromolecular confinement \cite{shmool2019observation}.  We observed that the confinement effect depends on the shape and size of the molecules. This effect cannot be observed in small organic molecular systems like sucrose where only few degrees of freedom of dihedral motion are available and hence it is not possible to reach a ``jammed conformation''. Bacitracin and lysozyme show a similar restriction of motions to that observed in HSA, i.e. a change of the PES. The effect is slightly more pronounced for lysozyme due to its increased size and hence higher number of internal degrees of freedom. Between \SIrange{310}{330}{K}, the absorption coefficient does not increase and the overall absorption change becomes less above \SI{300}{K}. At temperatures above \SI{330}{K}, the absorption increases again with temperature, as is also the case for BSA formulations measured previously \cite{shmool2019observation}. This temperature corresponds to an energy barrier of \SI{2.7}{kJ\,{mol^{-1}}} (equal to \SI{0.65}{kcal\,{mol^{-1}}}), which is significantly lower than the energy barrier of unfolding of BSA in solution, which is on the order of \SIrange{64}{267}{kJ/mol} \cite{NIKOLAIDIS2017235}. 

\begin{figure}
\centering

       \includegraphics[width=1\linewidth]{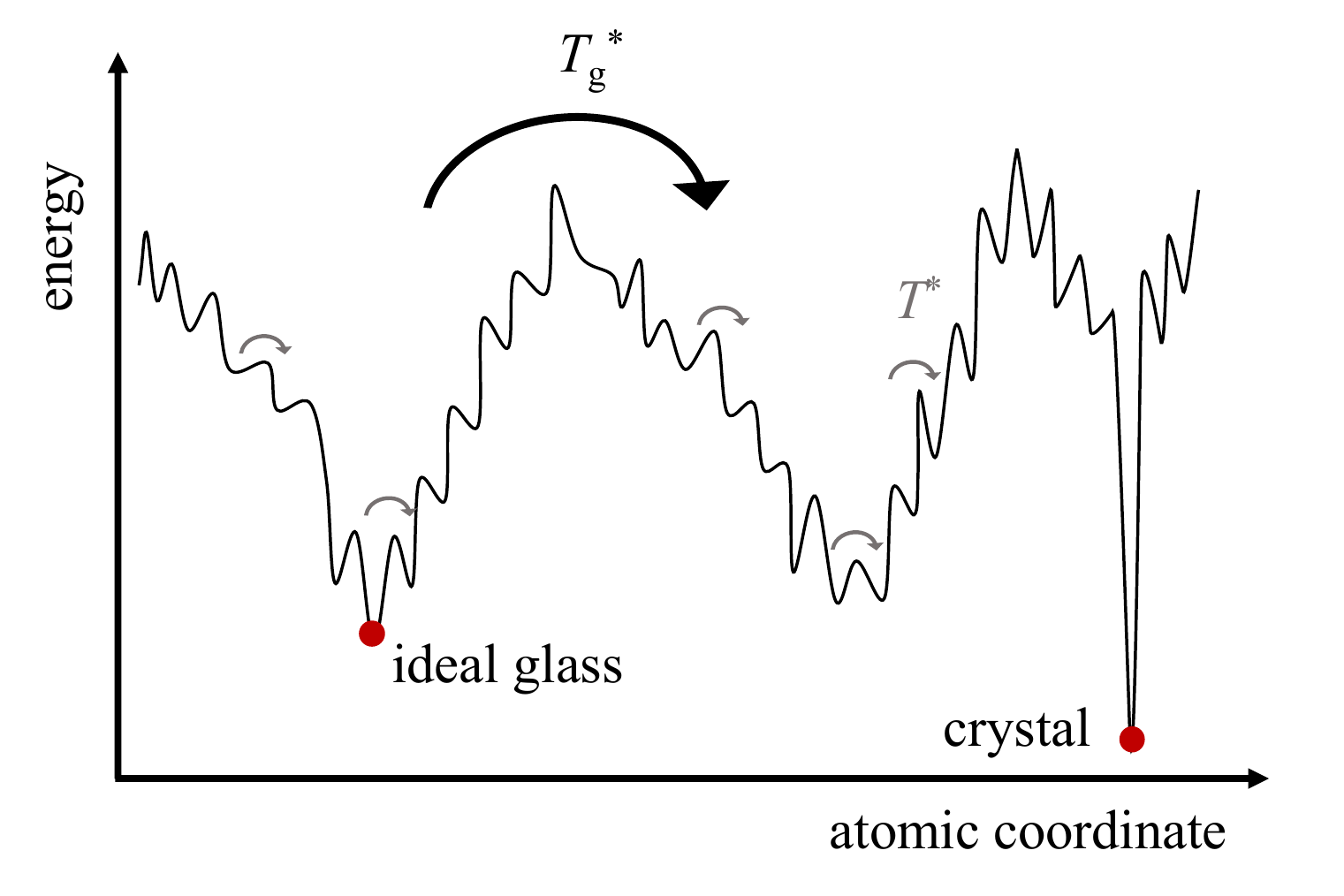}
	\caption{Schematic of a possible potential energy landscape topology; with sufficient thermal energy, a sample can explore different hypersurface configurations and can become trapped in shallow minima.  \label{fig:pes}}
\end{figure}

The confinement is most pronounced in HSA, the largest macromolecule studied. Here, the absorption coefficient even decreases slightly at temperatures between \SIrange{310}{360}{K}. This decrease in absorption could be because the molecules may lose some of the degrees of freedom that they already gained at lower temperatures as the conformational jamming increases once they become trapped in a steep minimum on the potential energy landscape, as shown schematically in Figure \ref{fig:pes}. Even after the jammed conformation is overcome, $\text{d}\alpha/\text{d}T$ in HSA is less than half compared to lysozyme.

\subsubsection{The low-frequency (boson peak) region}
\begin{figure}
\centering
         \includegraphics[width=1\linewidth]{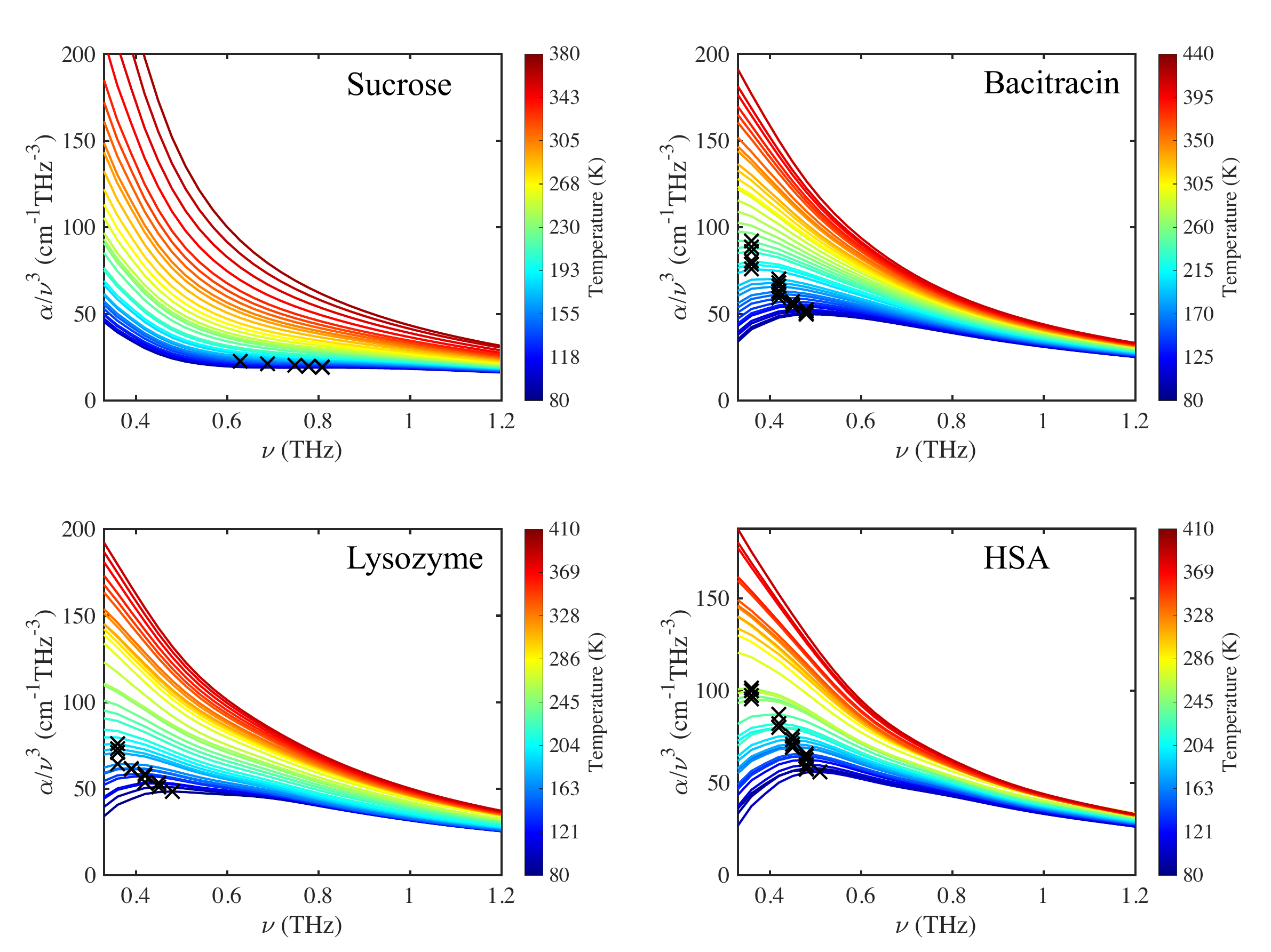}
	\caption{BP visualisation for sucrose, bacitracin, lysozyme, and HSA lyophilisates at different temperatures. Boson peaks have been highlighted with crosses. \label{fig:BP}}
\end{figure}

The BP can be visualised by plotting the absorption coefficient $\alpha$ divided by $\nu^3$ over the frequency $\nu$ (Figure \ref{fig:BP}),  \cite{kolbel2022terahertz} and the maximum value of $\alpha/\nu^3$ was found from smoothed data. 
If the BP occurs below \SI{0.3}{THz}, the data are not extrapolated, and no maximum is reported to avoid extrapolation errors.

In small molecular systems, for example, in glycerol, the dynamics of glasses upon heating usually fall into two regimes \cite{chumakov2004collective}. At very low temperatures, the terahertz spectra are dominated by harmonic excitations. The BP itself is harmonic and, therefore, a temperature-independent phenomenon. Its centre frequency is constant.
Above \Tgstar, the system leaves the harmonic minimum on the potential energy landscape. Shallower minima increase the amount of anharmonicity and absorption. Once anharmonic effects dominate at \Tstar, they lead to an apparent frequency shift of the BP maximum and obscure it completely at temperatures close to \Tgstar  \cite{kolbel2022terahertz}.

The maximum frequency of the BP of the protein lyophilisates is preserved in the harmonic regime at temperatures below approximately \SIrange{150}{200}{K}. However, it decreases in the anharmonic regime with increasing temperature before being obscured by anharmonic effects that appear to shift it outside the experimentally accessible region. 
The BP in sucrose is the least pronounced of the samples measured and appears to be masked by anharmonic effects already at very low temperatures.

The BP occurs close to the Ioffe-Regel crossover frequency at which the mean-free path of transverse waves becomes equal to their wavelength, meaning that there is a crossover from wave-like to random-matrix-like physics. Once global mobility sets in above the glass transition temperature, anharmonicity and mobility increase with temperature until a ``critical'' mobility is reached, completely obscuring the BP.  

At very similar temperatures, the proteins reconfigure and get trapped in a different conformation, decreasing mobility overall. 
Interestingly, the inflection point observed in the DSC data coincides with the temperature regime just above the initial trapping. Therefore, it is hypothesised that the trapping and/or the conformational change inducing the trapping result in a subtle heat capacity change with increasing temperature and that the maximum in the DSC data corresponds to partial unfolding.

\subsubsection{Anharmonicity parameter $a$}

The derivative parameter $a$ can be utilisied to characterise terahertz spectra at frequencies above and below the BP more reliably. It reflects the slope of the absorption spectra averaged over a specific frequency range \cite{kolbel2022terahertz}.

We chose three different frequency ranges: \SIrange{0.35}{0.55}{THz} (below the BP), \SIrange{0.90}{1.10}{THz} (above the BP at the spectrometer's highest SNR), and \SIrange{1.45}{1.65}{THz} (at even higher frequencies above the BP).

\begin{figure}
\centering
      \includegraphics[width=\linewidth]{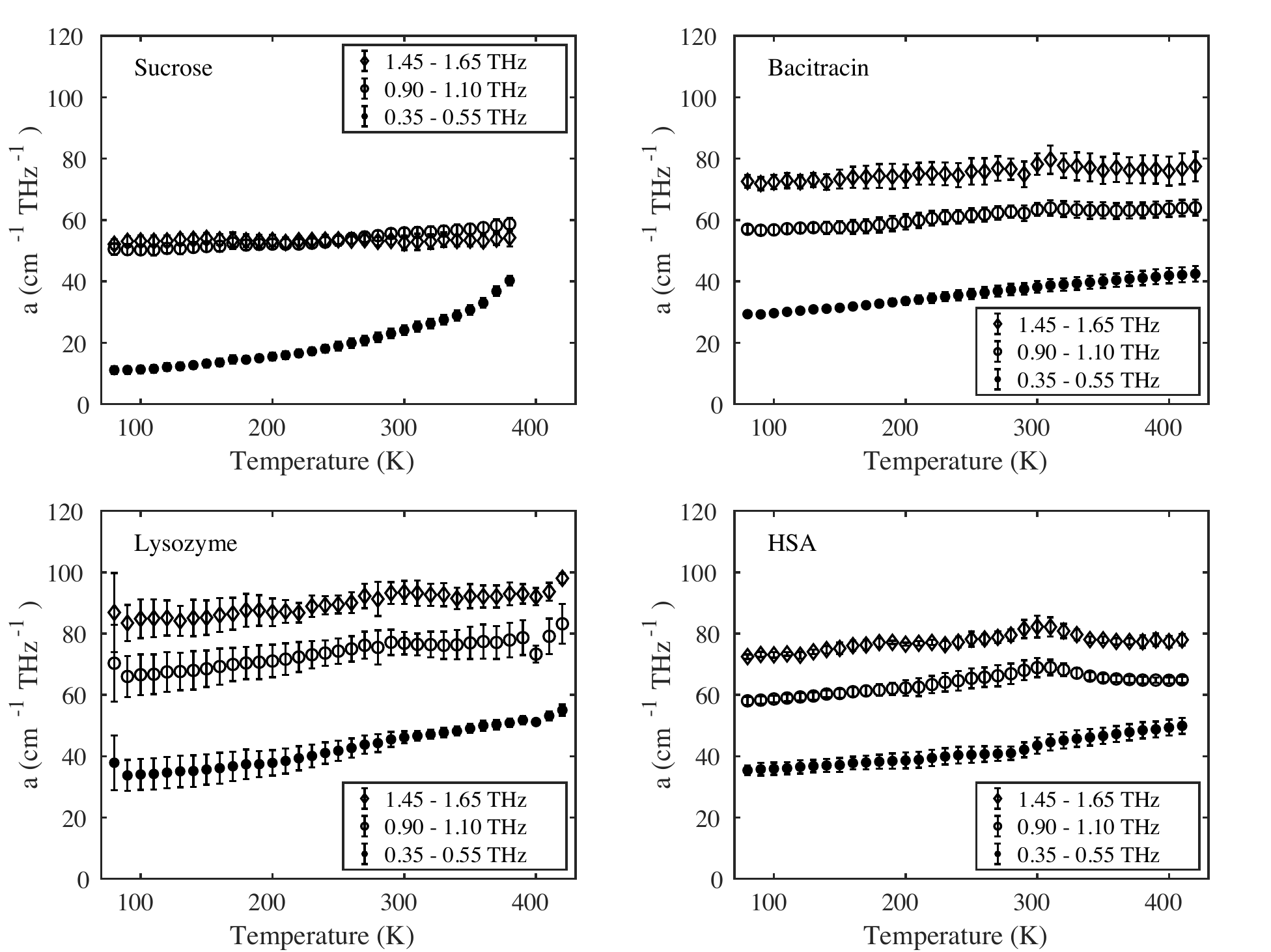}
	\caption{Anharmonicity parameter evaluated in different frequency ranges for sucrose, bacitracin, lysozyme, and HSA lyophilisates. Error bars are standard error for $n$ measurements ($n=5$ for sucrose, $n=4$ for bacitracin, $n=3$ for lysozyme and HSA). \label{fig:anharm_all_freq}}
\end{figure}

The lyophilisates show a markedly different behaviour at frequencies below and above the BP (Figure \ref{fig:anharm_all_freq}). In the protein samples, the plateau just above room temperature, seen in the absorption coefficient at \SI{1}{THz} (Figure \ref{fig:absat1_stacked}), can only be observed at frequencies above the BP. 
At lower frequencies, the increase of $a$ with temperature is monotonous. 
For bacitracin, that increase is approximately linear with temperature, while for lysozyme and HSA, a subtle transition can be observed at around \SI{200}{K} and \SI{300}{K}, respectively.

While the protein dynamical transition has thus far only been observed in hydrated samples, these increases in $a$ point to the existence of thermally activated modes that involve only the protein molecules themselves. Even without solvent molecules, parts of the proteins retain some flexibility. 

Given that the increase in $a$ appears in a similar temperature range, for example, in lysozyme, it could be caused by side chain motions where the activation energy is not affected by the presence of the solvent  \cite{markelz2007protein}.

Sucrose shows a more pronounced increase of $a$ over temperature, and a distinct increase is observed at \Tgstar. This discontinuity is expected as the glass transition temperature marks the onset of global mobility.

\subsubsection{Molecular confinement is observed in the spectrum at frequencies beyond the BP}

While the absorption coefficient at \SI{1}{THz} is located relatively close to the BP maximum and is strongly influenced by anharmonic effects, a different behaviour may be observed at higher frequencies. 

In the following we use a model developed by Chumakov et al. \cite{chumakov2004collective} to investigate the effect of subtle spectral changes on the extrapolated vibrational density of states (VDOS). This model was previously utilised to investigate the model glass former glycerol and is now applied to more complex biomolecules.

The exponential function $ \alpha = A \nu^3 \exp{(-\nu/\nu_c)}+C$ is fitted to the higher frequency part of the experimentally accessible spectrum. If that function is plotted at even higher frequencies, a peak with a centre frequency of $3\nu_c$ is observed. It has to be noted that this is a feature based on the fitting function and not an accurate prediction of the centre frequency of the VDOS. The experimental data available spans a frequency range from the Ioffe-Regel crossover up to approximately \SI{2.3}{THz}, while the actual VDOS may exhibit an underlying multi-peak structure \cite{verrall1988structure} and show more features beyond the peak itself \cite{ruggiero2016influence}.

Chumakov et al. found excellent agreement in the frequency range of \SIrange{1}{1.7}{THz} in glycerol, where an exponential decay in the reduced density of states was observed
 \cite{chumakov2004collective}. The reduced density of states and the absorption coefficient measured with THz-TDS are related by a factor of $\nu^3$.

\begin{figure}
\centering
      \includegraphics[width=\linewidth]{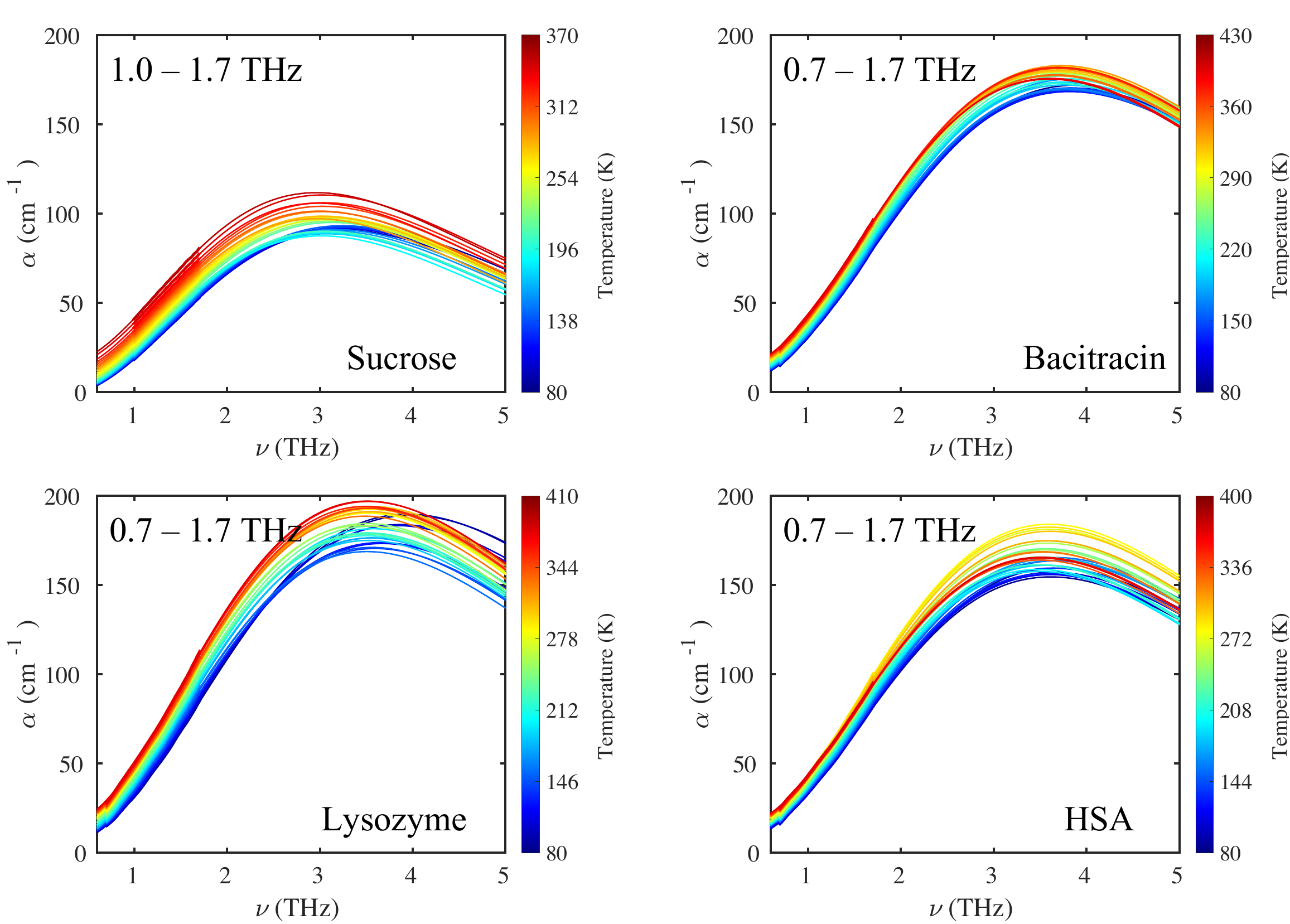}
	\caption{Extrapolated spectra at higher frequencies, for sucrose, bacitracin, lysozyme, and HSA lyophilisates, fit to frequencies above the BP maximum.  \label{fig:VDOS_extrapolated}}
\end{figure}

\begin{figure}
\centering
      \includegraphics[width=1\linewidth]{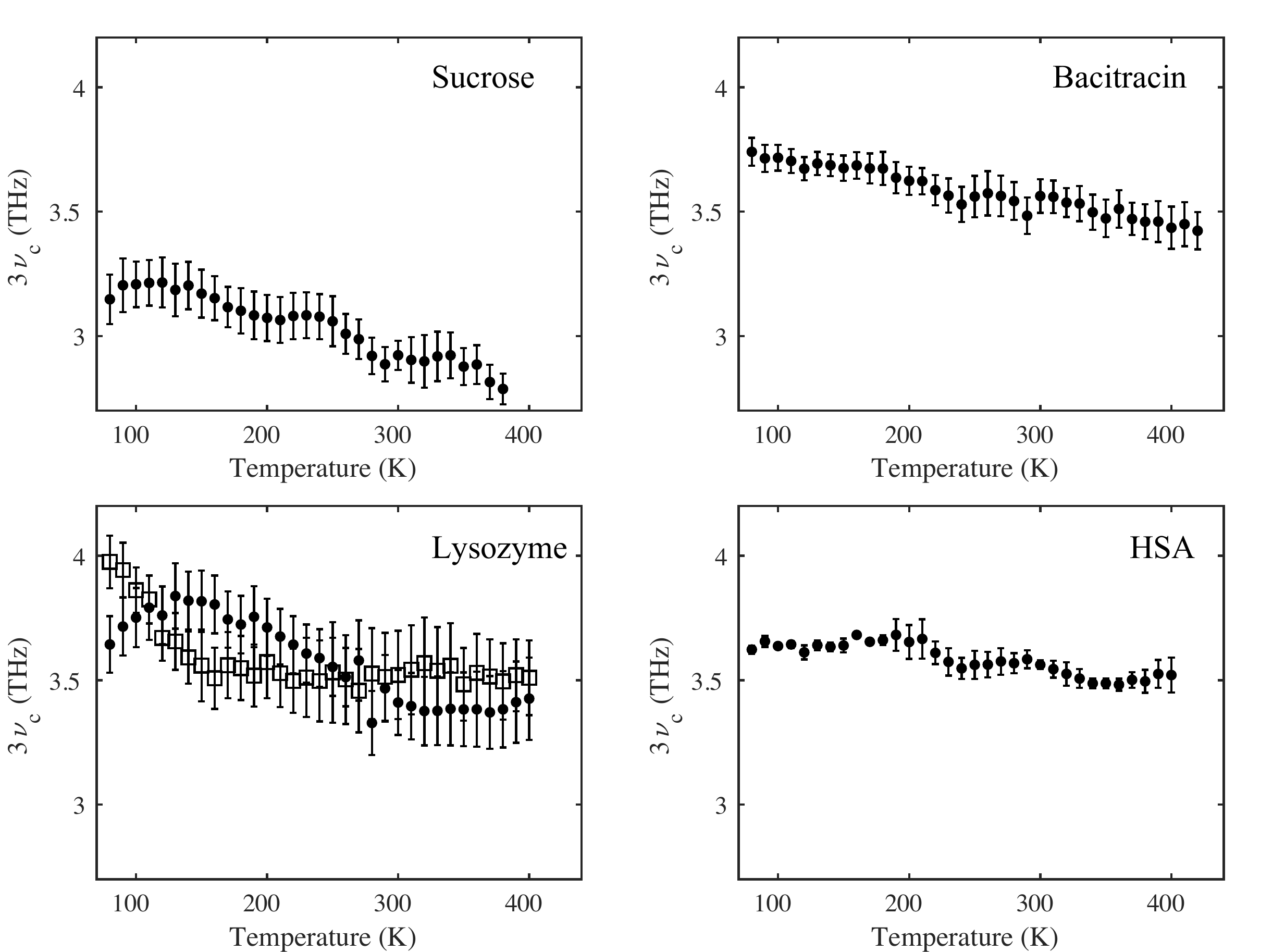}

	\caption{Extrapolated centre frequency for sucrose, bacitracin, lysozyme, and HSA lyophilisates. 
	Error bars are standard error for $n$ measurements ($n=5$ for sucrose, $n=4$ for bacitracin, $n=3$ for HSA). For the lysozyme lyophilisate, two separate measurements are shown with error bars the respective \SI{95}{\percent} confidence intervals of the fit.
	\label{fig:fc}}
\end{figure}

The peak described by this model is narrowest and least intense in sucrose and gets broader but more intense in the protein samples (Figure \ref{fig:VDOS_extrapolated}). Some temperature dependence of the centre frequency is also apparent (Figure \ref{fig:fc}).
In all samples, the centre frequency of the VDOS decreases with increasing temperature, thereby shifting the VDOS to lower frequencies and increasing the absorption coefficient measured at the shoulder (e.g. at \SI{1}{THz}). It is possible that the frequency shift may follow a Bose-Einstein distribution as previously observed for crystalline modes where thermal excitation was mediated by phonons populating an anharmonic potential. A redshift of a mode is observed when phonons are excited by sufficient thermal energy \cite{allen2021anharmonicity}.
However, because the data are only extrapolated, we refrain from fitting a model and will simply discuss it in broader, qualitative terms. In future, it might be beneficial to measure similar samples on a spectrometer with higher spectral bandwidth to be able to extract more accurate data.

For sucrose, the change in the centre frequency is pronounced above \Tgstar, whereas in bacitracin, the decrease is gradual over the entire temperature range but slightly increased between \SI{180}{K} and \SI{350}{K}. The centre frequency of lysozyme generally decreases with temperature below \SI{300}{K} and is constant above.
The centre frequency of HSA lyophilisates is constant upon heating to a temperature of \SI{210}{K}, followed by a decrease in $\nu_\text{BP}$ and reaching a minimum at \SI{340}{K}, followed by a slight increase. In all cases, the change in the centre frequency is less at temperatures at which confinement is observed, indicating that the VDOS does not shift. If the molecules are trapped in a conformation, one may assume that the vibrations are temperature independent.

The higher frequencies seem to be more affected by the activation of modes at around \SI{200}{K}. Modes at higher frequencies typically involve a lower reduced mass than low-frequency vibrations. The effect on the higher frequency modes, therefore, is an indication that the modes that become active at around \SI{200}{K} may involve side chains or functional groups rather than the heavy protein backbone, which would influence the lower frequencies instead.
In this respect, using a frequency of \SI{1}{THz} to analyse the systems is beneficial as it provides insight into anharmonic effects, disappearance of the BP at low temperatures, and activation of modes as well as jamming at increased temperatures.

\subsection{Fr{\"o}hlich condensates}

Fr{\"o}hlich postulated the existence of coherent vibrations in the gigahertz to terahertz range that help facilitate the biological function of biomolecules \cite{frohlich1977biological,frohlich1970long}.
In terahertz spectra, such coherent modes appear as peaks in the absorption spectrum that sharpen when the temperature is decreased. Simultaneously, the absorption coefficient at neighbouring frequencies decreases \cite{li2022situ}. No such features were found in any measurement (Figure \ref{fig:spectra})  
Our data hence show no evidence for Fr{\"o}hlich coherence or the existence of a Fr{\"o}hlich condensate or other quantum effects. 
In protein solution, terahertz vibrations are propagated by the solvent and coupled to dielectric relaxations. This may lead to the fast dissipation of any postulated coherence or localised states due to the similarity in their frequencies. In this case, only extraordinarily rigid or well-ordered molecular structures would be observable.
However, we do not find such evidence for inherent coherent states for the four molecules under investigation.

\section{Conclusions}

The importance of solvents and the solvation shell surrounding proteins for their function is widely recognized. In the present work, we tried to obtain insights into protein-protein interactions in the dry state analysing lyophilisates of sucrose, bacitracin, lysozyme, and HSA by terahertz spectroscopy.

The glass transition temperature, \Tg, was identified in sucrose by DSC, whereas \Tg\ could not be identified for bacitracin, lysozyme, and HSA. However, THz-TDS demonstrated an increase in mobility with temperature. 
An increase in temperature led to the activation of modes involving only the protein molecules, resulting in an increase in the absolute absorption coefficient and the anharmonicity parameter. The anharmonicity parameter showed a markedly different behaviour below and above the BP centre frequency. 
Utilising the theoretical model by Chumakov et al., the higher frequencies were also evaluated, and a redshift of the VDOS was predicted.
Anharmonicity began to influence the spectra in sucrose at \Tstar\ and resulted in an apparent shift of the centre frequency of the BP. A further increase in temperature and, thereby, mobility, led to the dissipation of the BP.
For the larger proteins lysozyme and HSA, jamming was observed at increased temperatures after the dissipation of the BP and could only be overcome by a further increase in temperature. Future experiments making use of a higher- bandwidth spectrometer can investigate the impact of temperature change on the VDOS.
In the frequency range investigated, we could not find evidence for the occurrence of Fr{\"o}hlich coherence.

\begin{acknowledgements}
All authors would like to thank Walter Schirmacher for insightful discussions about the nature of the boson peak and theoretical aspects. JK thanks the EPSRC Cambridge Centre for Doctoral Training in Sensor Technologies and Applications (EP/L015889/1) and AstraZeneca for funding. MLA would like to thank Erasmus+ for funding. All authors would like to thank the Cambridge-LMU Strategic Partnership for funding.  For the purpose of open access, the authors have applied a Creative Commons Attribution (CC BY) licence to any Author Accepted Manuscript version arising from this submission.
\end{acknowledgements}


\begin{thebibliography}{35}%
\makeatletter
\providecommand \@ifxundefined [1]{%
 \@ifx{#1\undefined}
}%
\providecommand \@ifnum [1]{%
 \ifnum #1\expandafter \@firstoftwo
 \else \expandafter \@secondoftwo
 \fi
}%
\providecommand \@ifx [1]{%
 \ifx #1\expandafter \@firstoftwo
 \else \expandafter \@secondoftwo
 \fi
}%
\providecommand \natexlab [1]{#1}%
\providecommand \enquote  [1]{``#1''}%
\providecommand \bibnamefont  [1]{#1}%
\providecommand \bibfnamefont [1]{#1}%
\providecommand \citenamefont [1]{#1}%
\providecommand \href@noop [0]{\@secondoftwo}%
\providecommand \href [0]{\begingroup \@sanitize@url \@href}%
\providecommand \@href[1]{\@@startlink{#1}\@@href}%
\providecommand \@@href[1]{\endgroup#1\@@endlink}%
\providecommand \@sanitize@url [0]{\catcode `\\12\catcode `\$12\catcode `\&12\catcode `\#12\catcode `\^12\catcode `\_12\catcode `\%12\relax}%
\providecommand \@@startlink[1]{}%
\providecommand \@@endlink[0]{}%
\providecommand \url  [0]{\begingroup\@sanitize@url \@url }%
\providecommand \@url [1]{\endgroup\@href {#1}{\urlprefix }}%
\providecommand \urlprefix  [0]{URL }%
\providecommand \Eprint [0]{\href }%
\providecommand \doibase [0]{https://doi.org/}%
\providecommand \selectlanguage [0]{\@gobble}%
\providecommand \bibinfo  [0]{\@secondoftwo}%
\providecommand \bibfield  [0]{\@secondoftwo}%
\providecommand \translation [1]{[#1]}%
\providecommand \BibitemOpen [0]{}%
\providecommand \bibitemStop [0]{}%
\providecommand \bibitemNoStop [0]{.\EOS\space}%
\providecommand \EOS [0]{\spacefactor3000\relax}%
\providecommand \BibitemShut  [1]{\csname bibitem#1\endcsname}%
\let\auto@bib@innerbib\@empty
\bibitem [{\citenamefont {Aoki}\ \emph {et~al.}(2016)\citenamefont {Aoki}, \citenamefont {Shiraki},\ and\ \citenamefont {Hattori}}]{aoki2016salt}%
  \BibitemOpen
  \bibfield  {author} {\bibinfo {author} {\bibfnamefont {K.}~\bibnamefont {Aoki}}, \bibinfo {author} {\bibfnamefont {K.}~\bibnamefont {Shiraki}},\ and\ \bibinfo {author} {\bibfnamefont {T.}~\bibnamefont {Hattori}},\ }\bibfield  {title} {\bibinfo {title} {Salt effects on the picosecond dynamics of lysozyme hydration water investigated by terahertz time-domain spectroscopy and an insight into the {H}ofmeister series for protein stability and solubility},\ }\href@noop {} {\bibfield  {journal} {\bibinfo  {journal} {Physical Chemistry Chemical Physics}\ }\textbf {\bibinfo {volume} {18}},\ \bibinfo {pages} {15060} (\bibinfo {year} {2016})}\BibitemShut {NoStop}%
\bibitem [{\citenamefont {Dahanayake}\ and\ \citenamefont {Mitchell-Koch}(2018)}]{dahanayake2018does}%
  \BibitemOpen
  \bibfield  {author} {\bibinfo {author} {\bibfnamefont {J.~N.}\ \bibnamefont {Dahanayake}}\ and\ \bibinfo {author} {\bibfnamefont {K.~R.}\ \bibnamefont {Mitchell-Koch}},\ }\bibfield  {title} {\bibinfo {title} {How does solvation layer mobility affect protein structural dynamics?},\ }\href@noop {} {\bibfield  {journal} {\bibinfo  {journal} {Frontiers in Molecular Biosciences}\ }\textbf {\bibinfo {volume} {5}},\ \bibinfo {pages} {65} (\bibinfo {year} {2018})}\BibitemShut {NoStop}%
\bibitem [{\citenamefont {Roeters}\ \emph {et~al.}(2017)\citenamefont {Roeters}, \citenamefont {Iyer}, \citenamefont {Pletikapi{\'c}}, \citenamefont {Kogan}, \citenamefont {Subramaniam},\ and\ \citenamefont {Woutersen}}]{roeters2017evidence}%
  \BibitemOpen
  \bibfield  {author} {\bibinfo {author} {\bibfnamefont {S.~J.}\ \bibnamefont {Roeters}}, \bibinfo {author} {\bibfnamefont {A.}~\bibnamefont {Iyer}}, \bibinfo {author} {\bibfnamefont {G.}~\bibnamefont {Pletikapi{\'c}}}, \bibinfo {author} {\bibfnamefont {V.}~\bibnamefont {Kogan}}, \bibinfo {author} {\bibfnamefont {V.}~\bibnamefont {Subramaniam}},\ and\ \bibinfo {author} {\bibfnamefont {S.}~\bibnamefont {Woutersen}},\ }\bibfield  {title} {\bibinfo {title} {Evidence for intramolecular antiparallel beta-sheet structure in alpha-synuclein fibrils from a combination of two-dimensional infrared spectroscopy and atomic force microscopy},\ }\href@noop {} {\bibfield  {journal} {\bibinfo  {journal} {Scientific reports}\ }\textbf {\bibinfo {volume} {7}},\ \bibinfo {pages} {41051} (\bibinfo {year} {2017})}\BibitemShut {NoStop}%
\bibitem [{\citenamefont {Galvagnion}\ \emph {et~al.}(2015)\citenamefont {Galvagnion}, \citenamefont {Buell}, \citenamefont {Meisl}, \citenamefont {Michaels}, \citenamefont {Vendruscolo}, \citenamefont {Knowles},\ and\ \citenamefont {Dobson}}]{galvagnion2015lipid}%
  \BibitemOpen
  \bibfield  {author} {\bibinfo {author} {\bibfnamefont {C.}~\bibnamefont {Galvagnion}}, \bibinfo {author} {\bibfnamefont {A.~K.}\ \bibnamefont {Buell}}, \bibinfo {author} {\bibfnamefont {G.}~\bibnamefont {Meisl}}, \bibinfo {author} {\bibfnamefont {T.~C.}\ \bibnamefont {Michaels}}, \bibinfo {author} {\bibfnamefont {M.}~\bibnamefont {Vendruscolo}}, \bibinfo {author} {\bibfnamefont {T.~P.}\ \bibnamefont {Knowles}},\ and\ \bibinfo {author} {\bibfnamefont {C.~M.}\ \bibnamefont {Dobson}},\ }\bibfield  {title} {\bibinfo {title} {Lipid vesicles trigger $\alpha$-synuclein aggregation by stimulating primary nucleation},\ }\href@noop {} {\bibfield  {journal} {\bibinfo  {journal} {Nature chemical biology}\ }\textbf {\bibinfo {volume} {11}},\ \bibinfo {pages} {229} (\bibinfo {year} {2015})}\BibitemShut {NoStop}%
\bibitem [{\citenamefont {Munishkina}\ \emph {et~al.}(2008)\citenamefont {Munishkina}, \citenamefont {Fink},\ and\ \citenamefont {Uversky}}]{munishkina2008concerted}%
  \BibitemOpen
  \bibfield  {author} {\bibinfo {author} {\bibfnamefont {L.~A.}\ \bibnamefont {Munishkina}}, \bibinfo {author} {\bibfnamefont {A.~L.}\ \bibnamefont {Fink}},\ and\ \bibinfo {author} {\bibfnamefont {V.~N.}\ \bibnamefont {Uversky}},\ }\bibfield  {title} {\bibinfo {title} {Concerted action of metals and macromolecular crowding on the fibrillation of $\alpha$-synuclein},\ }\href@noop {} {\bibfield  {journal} {\bibinfo  {journal} {Protein and peptide letters}\ }\textbf {\bibinfo {volume} {15}},\ \bibinfo {pages} {1079} (\bibinfo {year} {2008})}\BibitemShut {NoStop}%
\bibitem [{\citenamefont {Binolfi}\ \emph {et~al.}(2006)\citenamefont {Binolfi}, \citenamefont {Rasia}, \citenamefont {Bertoncini}, \citenamefont {Ceolin}, \citenamefont {Zweckstetter}, \citenamefont {Griesinger}, \citenamefont {Jovin},\ and\ \citenamefont {Fern{\'a}ndez}}]{binolfi2006interaction}%
  \BibitemOpen
  \bibfield  {author} {\bibinfo {author} {\bibfnamefont {A.}~\bibnamefont {Binolfi}}, \bibinfo {author} {\bibfnamefont {R.~M.}\ \bibnamefont {Rasia}}, \bibinfo {author} {\bibfnamefont {C.~W.}\ \bibnamefont {Bertoncini}}, \bibinfo {author} {\bibfnamefont {M.}~\bibnamefont {Ceolin}}, \bibinfo {author} {\bibfnamefont {M.}~\bibnamefont {Zweckstetter}}, \bibinfo {author} {\bibfnamefont {C.}~\bibnamefont {Griesinger}}, \bibinfo {author} {\bibfnamefont {T.~M.}\ \bibnamefont {Jovin}},\ and\ \bibinfo {author} {\bibfnamefont {C.~O.}\ \bibnamefont {Fern{\'a}ndez}},\ }\bibfield  {title} {\bibinfo {title} {Interaction of $\alpha$-synuclein with divalent metal ions reveals key differences: A link between structure, binding specificity and fibrillation enhancement},\ }\href@noop {} {\bibfield  {journal} {\bibinfo  {journal} {Journal of the American Chemical Society}\ }\textbf {\bibinfo {volume} {128}},\ \bibinfo {pages} {9893} (\bibinfo {year} {2006})}\BibitemShut {NoStop}%
\bibitem [{\citenamefont {Uversky}\ \emph {et~al.}(2001)\citenamefont {Uversky}, \citenamefont {Li},\ and\ \citenamefont {Fink}}]{uversky2001evidence}%
  \BibitemOpen
  \bibfield  {author} {\bibinfo {author} {\bibfnamefont {V.~N.}\ \bibnamefont {Uversky}}, \bibinfo {author} {\bibfnamefont {J.}~\bibnamefont {Li}},\ and\ \bibinfo {author} {\bibfnamefont {A.~L.}\ \bibnamefont {Fink}},\ }\bibfield  {title} {\bibinfo {title} {Evidence for a partially folded intermediate in $\alpha$-synuclein fibril formation},\ }\href@noop {} {\bibfield  {journal} {\bibinfo  {journal} {Journal of Biological Chemistry}\ }\textbf {\bibinfo {volume} {276}},\ \bibinfo {pages} {10737} (\bibinfo {year} {2001})}\BibitemShut {NoStop}%
\bibitem [{\citenamefont {Giehm}\ and\ \citenamefont {Otzen}(2010)}]{giehm2010strategies}%
  \BibitemOpen
  \bibfield  {author} {\bibinfo {author} {\bibfnamefont {L.}~\bibnamefont {Giehm}}\ and\ \bibinfo {author} {\bibfnamefont {D.~E.}\ \bibnamefont {Otzen}},\ }\bibfield  {title} {\bibinfo {title} {Strategies to increase the reproducibility of protein fibrillization in plate reader assays},\ }\href@noop {} {\bibfield  {journal} {\bibinfo  {journal} {Analytical biochemistry}\ }\textbf {\bibinfo {volume} {400}},\ \bibinfo {pages} {270} (\bibinfo {year} {2010})}\BibitemShut {NoStop}%
\bibitem [{\citenamefont {Stephens}\ \emph {et~al.}(2022)\citenamefont {Stephens}, \citenamefont {K{\"o}lbel}, \citenamefont {Moons}, \citenamefont {Chung}, \citenamefont {Ruggiero}, \citenamefont {Mahmoudi}, \citenamefont {Shmool}, \citenamefont {McCoy}, \citenamefont {Nietlispach}, \citenamefont {Routh} \emph {et~al.}}]{stephens2022decreased}%
  \BibitemOpen
  \bibfield  {author} {\bibinfo {author} {\bibfnamefont {A.~D.}\ \bibnamefont {Stephens}}, \bibinfo {author} {\bibfnamefont {J.}~\bibnamefont {K{\"o}lbel}}, \bibinfo {author} {\bibfnamefont {R.}~\bibnamefont {Moons}}, \bibinfo {author} {\bibfnamefont {C.~W.}\ \bibnamefont {Chung}}, \bibinfo {author} {\bibfnamefont {M.~T.}\ \bibnamefont {Ruggiero}}, \bibinfo {author} {\bibfnamefont {N.}~\bibnamefont {Mahmoudi}}, \bibinfo {author} {\bibfnamefont {T.~A.}\ \bibnamefont {Shmool}}, \bibinfo {author} {\bibfnamefont {T.~M.}\ \bibnamefont {McCoy}}, \bibinfo {author} {\bibfnamefont {D.}~\bibnamefont {Nietlispach}}, \bibinfo {author} {\bibfnamefont {A.~F.}\ \bibnamefont {Routh}}, \emph {et~al.},\ }\bibfield  {title} {\bibinfo {title} {Decreased water mobility contributes to increased $\alpha$-synuclein aggregation},\ }\href@noop {} {\bibfield  {journal} {\bibinfo  {journal} {Angewandte Chemie International Edition}\ } (\bibinfo {year} {2022})}\BibitemShut {NoStop}%
\bibitem [{\citenamefont {Khodadadi}\ and\ \citenamefont {Sokolov}(2015)}]{khodadadi2015protein}%
  \BibitemOpen
  \bibfield  {author} {\bibinfo {author} {\bibfnamefont {S.}~\bibnamefont {Khodadadi}}\ and\ \bibinfo {author} {\bibfnamefont {A.}~\bibnamefont {Sokolov}},\ }\bibfield  {title} {\bibinfo {title} {Protein dynamics: from rattling in a cage to structural relaxation},\ }\href@noop {} {\bibfield  {journal} {\bibinfo  {journal} {Soft Matter}\ }\textbf {\bibinfo {volume} {11}},\ \bibinfo {pages} {4984} (\bibinfo {year} {2015})}\BibitemShut {NoStop}%
\bibitem [{\citenamefont {Janc}\ \emph {et~al.}(2018)\citenamefont {Janc}, \citenamefont {Luk{\v{s}}i{\v{c}}}, \citenamefont {Vlachy}, \citenamefont {Rigaud}, \citenamefont {Rollet}, \citenamefont {Korb}, \citenamefont {M{\'e}riguet},\ and\ \citenamefont {Malikova}}]{janc2018ion}%
  \BibitemOpen
  \bibfield  {author} {\bibinfo {author} {\bibfnamefont {T.}~\bibnamefont {Janc}}, \bibinfo {author} {\bibfnamefont {M.}~\bibnamefont {Luk{\v{s}}i{\v{c}}}}, \bibinfo {author} {\bibfnamefont {V.}~\bibnamefont {Vlachy}}, \bibinfo {author} {\bibfnamefont {B.}~\bibnamefont {Rigaud}}, \bibinfo {author} {\bibfnamefont {A.-L.}\ \bibnamefont {Rollet}}, \bibinfo {author} {\bibfnamefont {J.-P.}\ \bibnamefont {Korb}}, \bibinfo {author} {\bibfnamefont {G.}~\bibnamefont {M{\'e}riguet}},\ and\ \bibinfo {author} {\bibfnamefont {N.}~\bibnamefont {Malikova}},\ }\bibfield  {title} {\bibinfo {title} {Ion-specificity and surface water dynamics in protein solutions},\ }\href@noop {} {\bibfield  {journal} {\bibinfo  {journal} {Physical Chemistry Chemical Physics}\ }\textbf {\bibinfo {volume} {20}},\ \bibinfo {pages} {30340} (\bibinfo {year} {2018})}\BibitemShut {NoStop}%
\bibitem [{\citenamefont {Leitner}\ \emph {et~al.}(2008)\citenamefont {Leitner}, \citenamefont {Gruebele},\ and\ \citenamefont {Havenith}}]{leitner2008solvation}%
  \BibitemOpen
  \bibfield  {author} {\bibinfo {author} {\bibfnamefont {D.~M.}\ \bibnamefont {Leitner}}, \bibinfo {author} {\bibfnamefont {M.}~\bibnamefont {Gruebele}},\ and\ \bibinfo {author} {\bibfnamefont {M.}~\bibnamefont {Havenith}},\ }\bibfield  {title} {\bibinfo {title} {Solvation dynamics of biomolecules: modeling and terahertz experiments},\ }\href@noop {} {\bibfield  {journal} {\bibinfo  {journal} {HFSP journal}\ }\textbf {\bibinfo {volume} {2}},\ \bibinfo {pages} {314} (\bibinfo {year} {2008})}\BibitemShut {NoStop}%
\bibitem [{\citenamefont {Tang}\ and\ \citenamefont {Pikal}(2004)}]{tang2004design}%
  \BibitemOpen
  \bibfield  {author} {\bibinfo {author} {\bibfnamefont {X.~C.}\ \bibnamefont {Tang}}\ and\ \bibinfo {author} {\bibfnamefont {M.~J.}\ \bibnamefont {Pikal}},\ }\bibfield  {title} {\bibinfo {title} {Design of freeze-drying processes for pharmaceuticals: practical advice},\ }\href@noop {} {\bibfield  {journal} {\bibinfo  {journal} {Pharmaceutical research}\ }\textbf {\bibinfo {volume} {21}},\ \bibinfo {pages} {191} (\bibinfo {year} {2004})}\BibitemShut {NoStop}%
\bibitem [{\citenamefont {Cicerone}\ \emph {et~al.}(2015)\citenamefont {Cicerone}, \citenamefont {Pikal},\ and\ \citenamefont {Qian}}]{cicerone2015stabilization}%
  \BibitemOpen
  \bibfield  {author} {\bibinfo {author} {\bibfnamefont {M.~T.}\ \bibnamefont {Cicerone}}, \bibinfo {author} {\bibfnamefont {M.~J.}\ \bibnamefont {Pikal}},\ and\ \bibinfo {author} {\bibfnamefont {K.~K.}\ \bibnamefont {Qian}},\ }\bibfield  {title} {\bibinfo {title} {Stabilization of proteins in solid form},\ }\href@noop {} {\bibfield  {journal} {\bibinfo  {journal} {Advanced drug delivery reviews}\ }\textbf {\bibinfo {volume} {93}},\ \bibinfo {pages} {14} (\bibinfo {year} {2015})}\BibitemShut {NoStop}%
\bibitem [{\citenamefont {Knab}\ \emph {et~al.}(2006)\citenamefont {Knab}, \citenamefont {Chen},\ and\ \citenamefont {Markelz}}]{knab2006hydration}%
  \BibitemOpen
  \bibfield  {author} {\bibinfo {author} {\bibfnamefont {J.}~\bibnamefont {Knab}}, \bibinfo {author} {\bibfnamefont {J.-Y.}\ \bibnamefont {Chen}},\ and\ \bibinfo {author} {\bibfnamefont {A.}~\bibnamefont {Markelz}},\ }\bibfield  {title} {\bibinfo {title} {Hydration dependence of conformational dielectric relaxation of lysozyme},\ }\href@noop {} {\bibfield  {journal} {\bibinfo  {journal} {Biophysical journal}\ }\textbf {\bibinfo {volume} {90}},\ \bibinfo {pages} {2576} (\bibinfo {year} {2006})}\BibitemShut {NoStop}%
\bibitem [{\citenamefont {Sibik}\ \emph {et~al.}(2013)\citenamefont {Sibik}, \citenamefont {Shalaev},\ and\ \citenamefont {Zeitler}}]{sibik2013glassy}%
  \BibitemOpen
  \bibfield  {author} {\bibinfo {author} {\bibfnamefont {J.}~\bibnamefont {Sibik}}, \bibinfo {author} {\bibfnamefont {E.~Y.}\ \bibnamefont {Shalaev}},\ and\ \bibinfo {author} {\bibfnamefont {J.~A.}\ \bibnamefont {Zeitler}},\ }\bibfield  {title} {\bibinfo {title} {Glassy dynamics of sorbitol solutions at terahertz frequencies},\ }\href@noop {} {\bibfield  {journal} {\bibinfo  {journal} {Physical Chemistry Chemical Physics}\ }\textbf {\bibinfo {volume} {15}},\ \bibinfo {pages} {11931} (\bibinfo {year} {2013})}\BibitemShut {NoStop}%
\bibitem [{\citenamefont {He}\ \emph {et~al.}(2008)\citenamefont {He}, \citenamefont {Ku}, \citenamefont {Knab}, \citenamefont {Chen},\ and\ \citenamefont {Markelz}}]{he2008protein}%
  \BibitemOpen
  \bibfield  {author} {\bibinfo {author} {\bibfnamefont {Y.}~\bibnamefont {He}}, \bibinfo {author} {\bibfnamefont {P.~I.}\ \bibnamefont {Ku}}, \bibinfo {author} {\bibfnamefont {J.}~\bibnamefont {Knab}}, \bibinfo {author} {\bibfnamefont {J.}~\bibnamefont {Chen}},\ and\ \bibinfo {author} {\bibfnamefont {A.}~\bibnamefont {Markelz}},\ }\bibfield  {title} {\bibinfo {title} {Protein dynamical transition does not require protein structure},\ }\href@noop {} {\bibfield  {journal} {\bibinfo  {journal} {Physical Review Letters}\ }\textbf {\bibinfo {volume} {101}},\ \bibinfo {pages} {178103} (\bibinfo {year} {2008})}\BibitemShut {NoStop}%
\bibitem [{\citenamefont {Cavagna}(2009)}]{cavagna2009supercooled}%
  \BibitemOpen
  \bibfield  {author} {\bibinfo {author} {\bibfnamefont {A.}~\bibnamefont {Cavagna}},\ }\bibfield  {title} {\bibinfo {title} {Supercooled liquids for pedestrians},\ }\href@noop {} {\bibfield  {journal} {\bibinfo  {journal} {Physics Reports}\ }\textbf {\bibinfo {volume} {476}},\ \bibinfo {pages} {51} (\bibinfo {year} {2009})}\BibitemShut {NoStop}%
\bibitem [{\citenamefont {Goldstein}(1969)}]{goldstein1969viscous}%
  \BibitemOpen
  \bibfield  {author} {\bibinfo {author} {\bibfnamefont {M.}~\bibnamefont {Goldstein}},\ }\bibfield  {title} {\bibinfo {title} {Viscous liquids and the glass transition: a potential energy barrier picture},\ }\href@noop {} {\bibfield  {journal} {\bibinfo  {journal} {The Journal of Chemical Physics}\ }\textbf {\bibinfo {volume} {51}},\ \bibinfo {pages} {3728} (\bibinfo {year} {1969})}\BibitemShut {NoStop}%
\bibitem [{\citenamefont {Shmool}\ and\ \citenamefont {Zeitler}(2019)}]{shmool2019insights}%
  \BibitemOpen
  \bibfield  {author} {\bibinfo {author} {\bibfnamefont {T.~A.}\ \bibnamefont {Shmool}}\ and\ \bibinfo {author} {\bibfnamefont {J.~A.}\ \bibnamefont {Zeitler}},\ }\bibfield  {title} {\bibinfo {title} {Insights into the structural dynamics of poly lactic-co-glycolic acid at terahertz frequencies},\ }\href@noop {} {\bibfield  {journal} {\bibinfo  {journal} {Polymer Chemistry}\ }\textbf {\bibinfo {volume} {10}},\ \bibinfo {pages} {351} (\bibinfo {year} {2019})}\BibitemShut {NoStop}%
\bibitem [{\citenamefont {K{\"o}lbel}\ \emph {et~al.}(2023)\citenamefont {K{\"o}lbel}, \citenamefont {Schirmacher}, \citenamefont {Shalaev},\ and\ \citenamefont {Zeitler}}]{kolbel2022terahertz}%
  \BibitemOpen
  \bibfield  {author} {\bibinfo {author} {\bibfnamefont {J.}~\bibnamefont {K{\"o}lbel}}, \bibinfo {author} {\bibfnamefont {W.}~\bibnamefont {Schirmacher}}, \bibinfo {author} {\bibfnamefont {E.}~\bibnamefont {Shalaev}},\ and\ \bibinfo {author} {\bibfnamefont {J.~A.}\ \bibnamefont {Zeitler}},\ }\bibfield  {title} {\bibinfo {title} {Terahertz dynamics in the glycerol-water system},\ }\href@noop {} {\bibfield  {journal} {\bibinfo  {journal} {Physical Review B}\ }\textbf {\bibinfo {volume} {107}},\ \bibinfo {pages} {104203} (\bibinfo {year} {2023})}\BibitemShut {NoStop}%
\bibitem [{\citenamefont {Shmool}\ \emph {et~al.}(2019)\citenamefont {Shmool}, \citenamefont {Woodhams}, \citenamefont {Leutzsch}, \citenamefont {Stephens}, \citenamefont {Gaimann}, \citenamefont {Mantle}, \citenamefont {Schierle}, \citenamefont {van~der Walle},\ and\ \citenamefont {Zeitler}}]{shmool2019observation}%
  \BibitemOpen
  \bibfield  {author} {\bibinfo {author} {\bibfnamefont {T.~A.}\ \bibnamefont {Shmool}}, \bibinfo {author} {\bibfnamefont {P.}~\bibnamefont {Woodhams}}, \bibinfo {author} {\bibfnamefont {M.}~\bibnamefont {Leutzsch}}, \bibinfo {author} {\bibfnamefont {A.~D.}\ \bibnamefont {Stephens}}, \bibinfo {author} {\bibfnamefont {M.~U.}\ \bibnamefont {Gaimann}}, \bibinfo {author} {\bibfnamefont {M.~D.}\ \bibnamefont {Mantle}}, \bibinfo {author} {\bibfnamefont {G.~S.~K.}\ \bibnamefont {Schierle}}, \bibinfo {author} {\bibfnamefont {C.~F.}\ \bibnamefont {van~der Walle}},\ and\ \bibinfo {author} {\bibfnamefont {J.~A.}\ \bibnamefont {Zeitler}},\ }\bibfield  {title} {\bibinfo {title} {Observation of high-temperature macromolecular confinement in lyophilised protein formulations using terahertz spectroscopy},\ }\href@noop {} {\bibfield  {journal} {\bibinfo  {journal} {International journal of pharmaceutics: X}\ }\textbf {\bibinfo {volume} {1}},\ \bibinfo {pages} {100022} (\bibinfo {year} {2019})}\BibitemShut {NoStop}%
\bibitem [{\citenamefont {Schirmacher}\ \emph {et~al.}(1998)\citenamefont {Schirmacher}, \citenamefont {Diezemann},\ and\ \citenamefont {Ganter}}]{schirmacher1998harmonic}%
  \BibitemOpen
  \bibfield  {author} {\bibinfo {author} {\bibfnamefont {W.}~\bibnamefont {Schirmacher}}, \bibinfo {author} {\bibfnamefont {G.}~\bibnamefont {Diezemann}},\ and\ \bibinfo {author} {\bibfnamefont {C.}~\bibnamefont {Ganter}},\ }\bibfield  {title} {\bibinfo {title} {Harmonic vibrational excitations in disordered solids and the “boson peak”},\ }\href@noop {} {\bibfield  {journal} {\bibinfo  {journal} {Physical review letters}\ }\textbf {\bibinfo {volume} {81}},\ \bibinfo {pages} {136} (\bibinfo {year} {1998})}\BibitemShut {NoStop}%
\bibitem [{\citenamefont {Markelz}\ \emph {et~al.}(2007)\citenamefont {Markelz}, \citenamefont {Knab}, \citenamefont {Chen},\ and\ \citenamefont {He}}]{markelz2007protein}%
  \BibitemOpen
  \bibfield  {author} {\bibinfo {author} {\bibfnamefont {A.~G.}\ \bibnamefont {Markelz}}, \bibinfo {author} {\bibfnamefont {J.~R.}\ \bibnamefont {Knab}}, \bibinfo {author} {\bibfnamefont {J.~Y.}\ \bibnamefont {Chen}},\ and\ \bibinfo {author} {\bibfnamefont {Y.}~\bibnamefont {He}},\ }\bibfield  {title} {\bibinfo {title} {Protein dynamical transition in terahertz dielectric response},\ }\href@noop {} {\bibfield  {journal} {\bibinfo  {journal} {Chemical Physics Letters}\ }\textbf {\bibinfo {volume} {442}},\ \bibinfo {pages} {413} (\bibinfo {year} {2007})}\BibitemShut {NoStop}%
\bibitem [{\citenamefont {Duvillaret}\ \emph {et~al.}(1996)\citenamefont {Duvillaret}, \citenamefont {Garet},\ and\ \citenamefont {Coutaz}}]{duvillaret1996reliable}%
  \BibitemOpen
  \bibfield  {author} {\bibinfo {author} {\bibfnamefont {L.}~\bibnamefont {Duvillaret}}, \bibinfo {author} {\bibfnamefont {F.}~\bibnamefont {Garet}},\ and\ \bibinfo {author} {\bibfnamefont {J.-L.}\ \bibnamefont {Coutaz}},\ }\bibfield  {title} {\bibinfo {title} {A reliable method for extraction of material parameters in terahertz time-domain spectroscopy},\ }\href@noop {} {\bibfield  {journal} {\bibinfo  {journal} {IEEE Journal of selected topics in quantum electronics}\ }\textbf {\bibinfo {volume} {2}},\ \bibinfo {pages} {739} (\bibinfo {year} {1996})}\BibitemShut {NoStop}%
\bibitem [{\citenamefont {Chang}\ and\ \citenamefont {Pikal}(2009)}]{chang2009mechanisms}%
  \BibitemOpen
  \bibfield  {author} {\bibinfo {author} {\bibfnamefont {L.~L.}\ \bibnamefont {Chang}}\ and\ \bibinfo {author} {\bibfnamefont {M.~J.}\ \bibnamefont {Pikal}},\ }\bibfield  {title} {\bibinfo {title} {Mechanisms of protein stabilization in the solid state},\ }\href@noop {} {\bibfield  {journal} {\bibinfo  {journal} {Journal of pharmaceutical sciences}\ }\textbf {\bibinfo {volume} {98}},\ \bibinfo {pages} {2886} (\bibinfo {year} {2009})}\BibitemShut {NoStop}%
\bibitem [{\citenamefont {Roe}\ and\ \citenamefont {Labuza}(2005)}]{roe2005glass}%
  \BibitemOpen
  \bibfield  {author} {\bibinfo {author} {\bibfnamefont {K.}~\bibnamefont {Roe}}\ and\ \bibinfo {author} {\bibfnamefont {T.}~\bibnamefont {Labuza}},\ }\bibfield  {title} {\bibinfo {title} {Glass transition and crystallization of amorphous trehalose-sucrose mixtures},\ }\href@noop {} {\bibfield  {journal} {\bibinfo  {journal} {International journal of food properties}\ }\textbf {\bibinfo {volume} {8}},\ \bibinfo {pages} {559} (\bibinfo {year} {2005})}\BibitemShut {NoStop}%
\bibitem [{\citenamefont {Nikolaidis}\ and\ \citenamefont {Moschakis}(2017)}]{NIKOLAIDIS2017235}%
  \BibitemOpen
  \bibfield  {author} {\bibinfo {author} {\bibfnamefont {A.}~\bibnamefont {Nikolaidis}}\ and\ \bibinfo {author} {\bibfnamefont {T.}~\bibnamefont {Moschakis}},\ }\bibfield  {title} {\bibinfo {title} {Studying the denaturation of bovine serum albumin by a novel approach of difference-{UV} analysis},\ }\href {https://doi.org/https://doi.org/10.1016/j.foodchem.2016.07.133} {\bibfield  {journal} {\bibinfo  {journal} {Food Chemistry}\ }\textbf {\bibinfo {volume} {215}},\ \bibinfo {pages} {235} (\bibinfo {year} {2017})}\BibitemShut {NoStop}%
\bibitem [{\citenamefont {Chumakov}\ \emph {et~al.}(2004)\citenamefont {Chumakov}, \citenamefont {Sergueev}, \citenamefont {Van~Buerck}, \citenamefont {Schirmacher}, \citenamefont {Asthalter}, \citenamefont {Rueffer}, \citenamefont {Leupold},\ and\ \citenamefont {Petry}}]{chumakov2004collective}%
  \BibitemOpen
  \bibfield  {author} {\bibinfo {author} {\bibfnamefont {A.}~\bibnamefont {Chumakov}}, \bibinfo {author} {\bibfnamefont {I.}~\bibnamefont {Sergueev}}, \bibinfo {author} {\bibfnamefont {U.}~\bibnamefont {Van~Buerck}}, \bibinfo {author} {\bibfnamefont {W.}~\bibnamefont {Schirmacher}}, \bibinfo {author} {\bibfnamefont {T.}~\bibnamefont {Asthalter}}, \bibinfo {author} {\bibfnamefont {R.}~\bibnamefont {Rueffer}}, \bibinfo {author} {\bibfnamefont {O.}~\bibnamefont {Leupold}},\ and\ \bibinfo {author} {\bibfnamefont {W.}~\bibnamefont {Petry}},\ }\bibfield  {title} {\bibinfo {title} {Collective nature of the boson peak and universal transboson dynamics of glasses},\ }\href@noop {} {\bibfield  {journal} {\bibinfo  {journal} {Physical review letters}\ }\textbf {\bibinfo {volume} {92}},\ \bibinfo {pages} {245508} (\bibinfo {year} {2004})}\BibitemShut {NoStop}%
\bibitem [{\citenamefont {Verrall}\ \emph {et~al.}(1988)\citenamefont {Verrall}, \citenamefont {Gladden},\ and\ \citenamefont {Elliott}}]{verrall1988structure}%
  \BibitemOpen
  \bibfield  {author} {\bibinfo {author} {\bibfnamefont {D.}~\bibnamefont {Verrall}}, \bibinfo {author} {\bibfnamefont {L.}~\bibnamefont {Gladden}},\ and\ \bibinfo {author} {\bibfnamefont {S.}~\bibnamefont {Elliott}},\ }\bibfield  {title} {\bibinfo {title} {The structure of phosphorus selenide glasses},\ }\href@noop {} {\bibfield  {journal} {\bibinfo  {journal} {Journal of Non-Crystalline Solids}\ }\textbf {\bibinfo {volume} {106}},\ \bibinfo {pages} {47} (\bibinfo {year} {1988})}\BibitemShut {NoStop}%
\bibitem [{\citenamefont {Ruggiero}\ \emph {et~al.}(2016)\citenamefont {Ruggiero}, \citenamefont {Sibik},\ and\ \citenamefont {Zeitler}}]{ruggiero2016influence}%
  \BibitemOpen
  \bibfield  {author} {\bibinfo {author} {\bibfnamefont {M.~T.}\ \bibnamefont {Ruggiero}}, \bibinfo {author} {\bibfnamefont {J.}~\bibnamefont {Sibik}},\ and\ \bibinfo {author} {\bibfnamefont {J.~A.}\ \bibnamefont {Zeitler}},\ }\bibfield  {title} {\bibinfo {title} {The influence of intermolecular forces on the terahertz response of amorphous materials},\ }in\ \href@noop {} {\emph {\bibinfo {booktitle} {2016 41st International Conference on Infrared, Millimeter, and Terahertz waves (IRMMW-THz)}}}\ (\bibinfo {organization} {IEEE},\ \bibinfo {year} {2016})\ pp.\ \bibinfo {pages} {1--2}\BibitemShut {NoStop}%
\bibitem [{\citenamefont {Allen}\ \emph {et~al.}(2021)\citenamefont {Allen}, \citenamefont {Sanders}, \citenamefont {Horvat},\ and\ \citenamefont {Lewis}}]{allen2021anharmonicity}%
  \BibitemOpen
  \bibfield  {author} {\bibinfo {author} {\bibfnamefont {J.}~\bibnamefont {Allen}}, \bibinfo {author} {\bibfnamefont {T.}~\bibnamefont {Sanders}}, \bibinfo {author} {\bibfnamefont {J.}~\bibnamefont {Horvat}},\ and\ \bibinfo {author} {\bibfnamefont {R.}~\bibnamefont {Lewis}},\ }\bibfield  {title} {\bibinfo {title} {Anharmonicity-driven redshift and broadening of sharp terahertz features of $\alpha$-glycine single crystal from 20 {K} to 300 {K}: {T}heory and experiment},\ }\href@noop {} {\bibfield  {journal} {\bibinfo  {journal} {Spectrochimica Acta Part A: Molecular and Biomolecular Spectroscopy}\ }\textbf {\bibinfo {volume} {244}},\ \bibinfo {pages} {118635} (\bibinfo {year} {2021})}\BibitemShut {NoStop}%
\bibitem [{\citenamefont {Fr{\"o}hlich}(1977)}]{frohlich1977biological}%
  \BibitemOpen
  \bibfield  {author} {\bibinfo {author} {\bibfnamefont {H.}~\bibnamefont {Fr{\"o}hlich}},\ }\bibfield  {title} {\bibinfo {title} {Biological control through long range coherence},\ }in\ \href@noop {} {\emph {\bibinfo {booktitle} {Synergetics: A Workshop Proceedings of the International Workshop on Synergetics at Schloss Elmau, Bavaria, May 2--7, 1977}}}\ (\bibinfo {organization} {Springer},\ \bibinfo {year} {1977})\ pp.\ \bibinfo {pages} {241--246}\BibitemShut {NoStop}%
\bibitem [{\citenamefont {Fr{\"o}hlich}(1970)}]{frohlich1970long}%
  \BibitemOpen
  \bibfield  {author} {\bibinfo {author} {\bibfnamefont {H.}~\bibnamefont {Fr{\"o}hlich}},\ }\bibfield  {title} {\bibinfo {title} {Long range coherence and the action of enzymes},\ }\href@noop {} {\bibfield  {journal} {\bibinfo  {journal} {Nature}\ }\textbf {\bibinfo {volume} {228}},\ \bibinfo {pages} {1093} (\bibinfo {year} {1970})}\BibitemShut {NoStop}%
\bibitem [{\citenamefont {Li}\ \emph {et~al.}(2022)\citenamefont {Li}, \citenamefont {Kolbel}, \citenamefont {Davis}, \citenamefont {Korter}, \citenamefont {Bond}, \citenamefont {Threlfall},\ and\ \citenamefont {Zeitler}}]{li2022situ}%
  \BibitemOpen
  \bibfield  {author} {\bibinfo {author} {\bibfnamefont {Q.}~\bibnamefont {Li}}, \bibinfo {author} {\bibfnamefont {J.}~\bibnamefont {Kolbel}}, \bibinfo {author} {\bibfnamefont {M.~P.}\ \bibnamefont {Davis}}, \bibinfo {author} {\bibfnamefont {T.~M.}\ \bibnamefont {Korter}}, \bibinfo {author} {\bibfnamefont {A.~D.}\ \bibnamefont {Bond}}, \bibinfo {author} {\bibfnamefont {T.}~\bibnamefont {Threlfall}},\ and\ \bibinfo {author} {\bibfnamefont {J.~A.}\ \bibnamefont {Zeitler}},\ }\bibfield  {title} {\bibinfo {title} {In situ observation of the structure of crystallizing magnesium sulfate heptahydrate solutions with terahertz transmission spectroscopy},\ }\href@noop {} {\bibfield  {journal} {\bibinfo  {journal} {Crystal Growth \& Design}\ }\textbf {\bibinfo {volume} {22}},\ \bibinfo {pages} {3961} (\bibinfo {year} {2022})}\BibitemShut {NoStop}%
\end{thebibliography}
\end{document}